\documentclass[aps,twocolumn,prd,showpacs,showkeys,preprintnumbers,superscriptaddress,nobibnotes,floatfix,longbibliography]{revtex4-1}

\pdfoutput=1

\usepackage{graphicx}
\usepackage{bm}
\usepackage{times}
\usepackage{hyperref}
\usepackage{slashed}
\usepackage{color}
\usepackage{aas_macros}
\usepackage{nicefrac}
\usepackage{soul}
\usepackage{amstext}

\usepackage{slashed}
\usepackage{lipsum}
\usepackage{subfigure}
\usepackage{multirow}
\usepackage{amsmath}
\usepackage{array}
\usepackage{varwidth}
\usepackage{comment}

\hypersetup{
    pdfnewwindow=true,    
    colorlinks=true,      
    linkcolor=blue,       
    citecolor=blue,       
    filecolor=blue,      
    urlcolor=blue         
}
\newcommand{\orcid}[1]
{\begingroup
  \hypersetup{hidelinks}\href{https://orcid.org/#1}{\includegraphics[width=10pt]{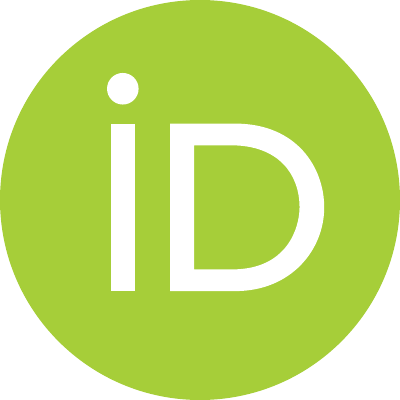}
} \endgroup}

\graphicspath{{figures/}}

\begin{document}

\title{Where are Milky Way's Hadronic PeVatrons?}

\author{Takahiro Sudoh \orcid{0000-0002-6884-1733}}
\email{sudo.4@osu.edu}
\affiliation{Center for Cosmology and AstroParticle Physics (CCAPP), Ohio State University, Columbus, OH 43210, USA}
\affiliation{Department of Physics, Ohio State University, Columbus, OH 43210, USA}
\affiliation{Department of Astronomy, Ohio State University, Columbus, OH 43210, USA}

\author{John F. Beacom \orcid{0000-0002-0005-2631}}
\email{beacom.7@osu.edu}
\affiliation{Center for Cosmology and AstroParticle Physics (CCAPP), Ohio State University, Columbus, OH 43210, USA}
\affiliation{Department of Physics, Ohio State University, Columbus, OH 43210, USA}
\affiliation{Department of Astronomy, Ohio State University, Columbus, OH 43210, USA}

\date{\today}

%%%%%%%%%%%%%%%%%%%%%%%%%%%%%%%%%%%%%%%%%%%%%%%%%%%%%%%%%%%%%%%%%%%%%%%%%%%%%%%%

\begin{abstract}

Observations of the Milky Way at TeV--PeV energies reveal a bright diffuse flux of hadronic cosmic rays and also bright point sources of gamma rays. If the gamma-ray sources are hadronic cosmic-ray accelerators, then they must also be neutrino sources.  However, no neutrino sources have been detected.  Where are they?  We introduce a new population-based approach to probe Milky Way hadronic PeVatrons, demanding consistency between diffuse and point-source PeV-range data on cosmic rays, gamma rays, and neutrinos.  For the PeVatrons, two extreme scenarios are allowed: (1) the hadronic cosmic-ray accelerators and the gamma-ray sources are the same objects, so that bright neutrino sources exist and improved telescopes can detect them, versus (2) the hadronic cosmic-ray accelerators and the gamma-ray sources are distinct, so that there are no detectable neutrino sources.  The latter case is possible if hadronic accelerators have sufficiently thin column densities.  We quantify present constraints and future prospects, showing how to reveal the nature of the hadronic PeVatrons.

\end{abstract}

\maketitle

%%%%%%%%%%%%%%%%%%%%%%%%%%%%%%%%%%%%%%%%%%%%%%%%%%%%%%%%%%%%%%%%%%%%%%%%%%%%%%%%
%%%%%%%%%%%%%%%%%%%%%%%%%%%%%%%%%%%%%%%%%%%%%%%%%%%%%%%%%%%%%%%%%%%%%%%%%%%%%%%%

\section{Introduction}

The Milky Way is blazing in the GeV--PeV diffuse emission of hadronic cosmic rays (CRs)~\cite{1964ocr..book.....G, 1990acr..book.....B, 1990cup..book.....G, 2016crpp.book.....G}.  This requires the existence of powerful accelerators, though their locations are obscured by magnetic deflections during CR propagation.  While many source classes seem able to accelerate hadrons to GeV energies, it remains mysterious which ones can reach PeV energies~\cite{1983A&A...125..249L, 2004MNRAS.353..550B, 2005APh....24....1A, 2013MNRAS.431..415B, 2013A&ARv..21...70B, 2014JHEAp...1....1A, 2021Univ....7..324C} --- i.e., the hadronic PeVatrons, which are our focus.  In principle, these should emit bright fluxes of gamma rays and neutrinos due to the pion-producing CR interactions with source matter and photons~\cite{1971NASSP.249.....S, 2004vhec.book.....A, 2009herb.book.....D}.

The Milky Way is also blazing in the GeV--PeV point-source emission of gamma rays~\cite{2007ApJ...664L..91A, 2017ApJ...843...40A, 2018A&A...612A...1H, 2020ApJS..247...33A, 2021Natur.594...33C}.  As the detector energy range is increased, sources become rarer, but they are still found, which indicates their powerful emission at high energies.  Recently, the Large High Altitude Air Shower Observatory (LHAASO) detected twelve northern-sky sources {above 100~TeV}~\cite{2021Natur.594...33C}.  It is often assumed that the highest-energy gamma-ray sources are hadronic CR accelerators. This may be true.  But it might instead be true that these sources are only leptonic CR accelerators, which produce gamma rays --- but not neutrinos --- through the inverse-Compton scattering of CR electrons with source photons~\cite{1996MNRAS.278..525A}.

\begin{figure}[b]
    \centering
    \includegraphics[width=\columnwidth]{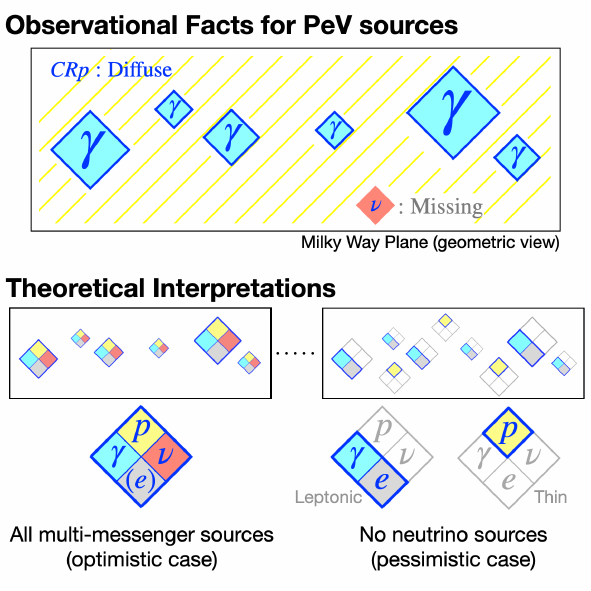}
    \caption{Possible scenarios for PeV sources.  Each diamond is a source, emitting the noted particles.  Bottom left shows the ``optimistic" case, where all gamma-ray sources are hadronic PeVatrons with neutrino emission.  Bottom right shows the ``pessimistic" case, where all gamma-ray sources are leptonic and no hadronic PeVatrons produce detectable gamma-ray or neutrino fluxes.}
    \label{fig:schematic}
\end{figure}

How can we identify the Milky Way's hadronic PeVatrons?  Neutrino emission would be a smoking gun.  However, no sources have been found in more than a decade of searches~\cite{2018PrPNP.102...73A, 2019ApJ...886...12A, 2020PhRvL.124e1103A, 2020ApJ...892...92A, 2020Ap&SS.365..108K, 2016ApJ...823...65A, 2017PhRvD..96h2001A} despite a series of optimistic predictions~\cite{2006PhRvD..74f3007K, 2007PhRvD..75h3001B, 2008PhRvD..78f3004H, 2009APh....31..437G}.  Figure~\ref{fig:schematic} schematically illustrates the range of possibilities.  In the optimistic case, the highest-energy gamma-ray sources are hadronic CR accelerators that produce neutrinos and electrons, making them exciting multi-messenger sources.  Then bright neutrino sources exist and will be found with future detectors~\cite{2021JPhG...48f0501A, 2016JPhG...43h4001A, 2018arXiv180810353B,2020NatAs...4..913A,  2022arXiv220704519Y}.  In the pessimistic case, the highest-energy gamma ray sources are all leptonic CR accelerators.  However, this seems to be at odds with the requirement that some sources must accelerate hadronic CRs.  Here we point out a viable possibility that has received little attention: the hadronic CR accelerators may be so \emph{thin} in column density that the CRs escape without {in-situ} interaction.  Then bright neutrino sources would not exist and identifying the hadronic CR accelerators would be difficult.

%%%%%%%%%%%%%%%%%%%%%%%%%%%%%%%%%%%%%%
%%%%%%%%%%%%%%%%%%%%%%%%%%%%%%%%%%%%%%

We distinguish the last possibility from so-called hidden sources, which are hadronic CR accelerators that are so \emph{thick} {in matter or radiation density} that gamma rays cannot escape, though neutrinos can.  But then these sources are not what we call hadronic PeVatrons, because in many cases the CRs cannot escape either, leaving unanswered where the Milky Way hadronic CRs come from.  Hidden sources are interesting, but they are not our focus.

In this paper, we present a new theoretical framework to systematically study Milky Way hadronic PeVatrons, introducing three innovations.  First, contrary to previous studies on individual candidate sources, here we take a population-studies approach.  Second, we also take a multi-messenger approach, demanding consistency with diffuse and point-source data on CRs, gamma rays, and neutrinos.  Third, we quantify the properties of PeVatrons in a semi-model-independent plane of source gas density ($n_{\rm src}^{\rm gas}$) and CR escape time ($\tau_{\rm src}^{\rm esc}$).  As the inputs are uncertain over orders of magnitude, we aim for a precision of a factor of a few, which we show is adequate given present flux sensitivities.  Our goal is to guide the next steps as observations and theory improve.  Our new theoretical framework is a valuable complement to other approaches.

In Sec.~\ref{sec:review}, we review very high energy astronomy.  In Sec.~\ref{sec:analytical-population}, we describe our models for the hadronic source population in a simplified way, followed in Sec.~\ref{sec:model-population} by the full details.  In Sec.~\ref{sec:result-population}, we calculate constraints based on current observations.  In Sec.~\ref{sec:result-population-future}, we calculate prospects for future neutrino telescopes and discuss ways to make progress.  In Sec.~\ref{sec:conclusion}, we summarize key new insights.

%%%%%%%%%%%%%%%%%%%%%%%%%%%%%%%%%%%%%%%%%%%%%%%%%%%%%%%%%%%%%%%%%%%%%%%%%%%%%%%%
%%%%%%%%%%%%%%%%%%%%%%%%%%%%%%%%%%%%%%%%%%%%%%%%%%%%%%%%%%%%%%%%%%%%%%%%%%%%%%%%

\section{Overview of Very High Energy Emission}
\label{sec:review}

In this section, we review observations and theory for very high energy astrophysics. We start with diffuse hadronic CR emission, then point sources of gamma rays and neutrinos, and finally diffuse gamma-ray and neutrino emission.

The CR spectrum is dominated by protons. The observed CR proton intensity~\cite{2009BRASP..73..564P, 2011Sci...332...69A, 2015PhRvL.114q1103A, 2017ApJ...839....5Y, 2018JETPL.108....5A, 2019SciA....5.3793A, 2019PhRvL.122r1102A} is
\begin{equation}
E_{p}^{2}\Phi_{p} \simeq 7\times 10^{-5}\left(\frac{E_{p}}{\rm PeV}\right)^{-0.7}{\rm GeV~cm^{-2}~s^{-1}~sr^{-1}}.
\label{eq:cr-obs-intensity}
\end{equation}
Near the PeV range, helium and other heavy components are also important~\cite{2006JPhCS..47...41H, 2022PhRvD.105f3021A}; their contribution to hadronic gamma-ray emission is discussed below.  The all-particle CR spectrum has a break at about 3~PeV, commonly defined as the ``knee"~\cite{2008ApJ...678.1165A, 2013APh....47...54A, 2019PhRvD.100h2002A, 2020APh...12002441L}; CRs below the knee are believed to be of a Galactic origin. {(Note that the knee is likely the onset of decrease in the proton spectrum; it does not necessarily correspond to the end of Galactic component.)} {Supernova remnants (SNRs) are} the most promising candidate up to the knee and possibly beyond~\cite{2010ApJ...718...31P}, although decisive conclusions are yet to be reached {due to the absence of observational proofs}~\cite{2017MNRAS.472.2956A, 2020ApJ...894...51A} {as well as some theoretical calculations showing maximum energy not reaching the knee~(e.g., Ref.~\cite{2022MNRAS.516..492B})}. For population approaches based on specific SNR models, see Refs.~\cite{2013MNRAS.434.2748C, 2018MNRAS.479.3415C}. We strive to make our constraints general, so that they also cover other scenarios~\cite{2014A&ARv..22...77B, 2017JCAP...04..037F, 2017MNRAS.470.3332I, 2018JCAP...07..042G, 2018MNRAS.478..926O, 2018ApJ...866...51K, 2020ApJ...889..146S, 2020MNRAS.493.3212C, 2020ApJ...904..188K, 2020A&A...635A.138G, 2021MNRAS.504.6096M, 2021ApJ...915...31K}.

An important observable is the grammage that CRs accumulate before escaping the Milky Way, which is constrained by data on the secondary-to-primary ratio~\cite{1990A&A...233...96E, 2013PhRvL.111u1101B}:
\begin{equation}
X \simeq 8.7~{\rm g~cm^{-2}}\left(\frac{E_p}{\rm 10~GeV}\right)^{-\delta}.
\label{eq:X-obs}
\end{equation}
Above 65 GeV/nucleon, observations are consistent with a power law of $\delta=0.33$ (at lower energies, the index is somewhat larger), but detailed constraints are available only below 3~TeV/nucleon, inducing uncertainties~\cite{2016PhRvL.117w1102A}. {Also note that the slope for the B/C ratio can differ from the slope for the grammage.} Nevertheless, we assume that the scaling $X \propto (E_p)^{-0.33}$ holds up to the knee, as is expected for CRs that scatter with external turbulence (as opposed to turbulence generated by CRs themselves)~\cite{2012PhRvL.109f1101B, 2013JCAP...07..001A}. {(Shortly below, we discuss uncertainties due to this assumption.)}

The Milky Way CR production rate can be estimated from the observed flux as
\begin{equation}
E_{p}^2 \frac{d^2N_{p}}{dE_{p} dt}\biggr\rvert_{\rm MW} = \frac{4\pi}{c} E_{p}^2\Phi_{p}\frac{V_{\rm CR}}{\tau^{\rm esc}(E_{p})},
\end{equation}
where $V_{\rm CR}$ is the Galactic volume that CRs are confined within and $\tau^{\rm esc}$ is the escape time from it.  (Note that here we focus on the Milky Way as a whole, whereas in most of the paper we focus on sources.) This equation is derived assuming steady-state conditions, in which case the CR energy density is the production rate on the left-hand side times $\tau^{\rm esc}/V_{\rm CR}$. In the standard paradigm, it is assumed that the grammage, at least in the GeV range, is dominated by the diffuse interstellar medium (ISM) \cite{1998ApJ...493..694M, 2007ARNPS..57..285S}, (see, however, Refs.~\cite{2010PhRvD..82b3009C, 2017PhRvD..95f3009L, 2019PhRvD..99d3005L, 2019IJMPD..2830022G}). Then, the factor $V_{\rm CR}/\tau^{\rm esc}$ is calculated via $X = \mu m_p n_{\rm ISM}^{\rm gas} c\tau^{\rm esc} = (M_{\rm gas}/V_{\rm CR}) c\tau^{\rm esc}$, where $m_p$ is the proton mass, $\mu\simeq1.4$ accounts for the composition of the {ISM} gas {on average}, $n_{\rm ISM}^{\rm gas}$ is the gas number density, and $M_{\rm gas}$ is the Milky Way's gas mass (set to $10^{10} M_\odot$~\cite{2019PhRvD..99f3012M}).  The energy-dependent CR production rate is then
\begin{equation}
\label{eq:CR_production}
E_{p}^2 \frac{d^2N_{p}}{dE_{p} dt}\biggr\rvert_{\rm MW} \sim 1.3\times 10^{38} \left(\frac{E_{p}}{\rm PeV}\right)^{-0.37}~{\rm erg~s^{-1}},
\end{equation}
which defines the energy budget of hadronic PeV sources~\cite{2019PhRvD..99f3012M}. {Integrated above 1~GeV, this yields the proton luminosity of $\sim6\times10^{40}$~erg~s$^{-1}$.} Compared to widely assumed {$dN/dE\propto E^{-2}$ or $E^{-2.2}$}, the above spectral index is softer,
yielding less power in the PeV range for a fixed total energy. {(An even softer $E^{-2.4}$ is also widely used \cite{2007ARNPS..57..285S}.)} We focus on the flux in a narrow energy range near 1~PeV, which makes our results robust to spectrum uncertainties. The CR spectrum from PeVatrons might be harder than -2.37; the observed power-law spectrum can be still obtained with additional soft-spectrum sources with lower maximum energies. The typical spectral index for PeVatrons is uncertain, but unimportant here, as our analysis is independent of the lower-energy emission.

The major uncertainties in Eq.~(\ref{eq:CR_production}) are the observed intensity and CR grammage. First, the observed proton intensities differ between experiments by a factor of about 1.6 at 1~PeV, which increases to 2.5 at 3~PeV~\cite{2020APh...12002441L}. {(See also Ref.~\cite{2015PhRvD..92i2005B}, which suggests the location of ``proton knee" might be lower than 3~PeV.)} Eq.~(\ref{eq:cr-obs-intensity}) uses the upper end of the range.  The actual production rate near 1 PeV might be smaller by a factor of 3. At higher energies, discrepancies between measurements are even larger (a factor of 6 at 10~PeV). Although such high energies are not our focus, in Sec.~\ref{subsec:gamma_etc} we comment on how future LHAASO gamma-ray data might help to resolve this tension. Second, we extrapolate $X$ from 3~TeV to 1~PeV assuming $\delta=0.33$, which is predicted by the Kolmogorov theory. Had we used $\delta=0.5$, as predicted by Kraichnan theory, the value of $X$ would be a factor of (1~PeV / 3~TeV)$^{0.17}$ smaller, which means that the actual production rate might be larger by a factor of 3. For the overall uncertainties on the production rate near $\sim$1~PeV, we thus expect a factor less than $\sim3$. 

At the sources or during propagation, CR hadrons collide with gas, producing roughly equal numbers of $\pi^+$, $\pi^-$, and $\pi^0$. The decay of a $\pi^0$ meson generates two gamma rays, each with energy $E_\gamma \sim 0.1 E_p$ {on average}, while the decays of $\pi^+$ and $\pi^-$ (and the subsequent decays of $\mu^+$ and $\mu^-$) produce three neutrinos, each with an energy of $E_\nu \sim 0.05E_p$ (see Refs.~\cite{2006PhRvD..74c4018K,  2014PhRvD..90l3014K, 2020ApJ...903...61C} for details). Hadronic gamma rays must thus be accompanied by neutrinos with comparable numbers and energies.

Gamma-ray observatories have detected a large number and large variety of Galactic sources in the GeV--PeV range. The search for the sources of $E_p\sim1$~PeV protons is most effective for secondaries around $E_\gamma\sim100~$TeV. Gamma-ray sources at such high energies are now detected. Tibet AS$\gamma$ and the High-Altitude Water Cherenkov Observatory (HAWC) have detected several gamma-ray sources emitting above 100~TeV~\cite{2019PhRvL.123e1101A, 2020PhRvL.124b1102A, 2020ApJ...896L..29A, 2021ApJ...907L..30A, 2021NatAs...5..460T, 2021PhRvL.127c1102A, 2021NatAs...5..465A,  2022ApJ...932..120A}. LHAASO has expanded the source count above 100~TeV, detecting twelve uniformly selected sources in Ref.~\cite{2021Natur.594...33C}, which we use.  (They also reported one more in  Ref.~\cite{2021ApJ...917L...4C}.) {Incredibly, some are observed at energies beyond 1 PeV.} These telescopes, combined with observations by imaging atmospheric Cherenkov telescopes (IACTs), have discovered particularly strong candidates for hadronic PeVatrons~\cite{2020ApJ...896L..29A, 2021ApJ...907L..30A, 2021NatAs...5..460T, 2021A&A...653A.152A, 2019ApJ...885..162X, 2022PhRvL.129g1101F}.

On the contrary, no Galactic neutrino sources have been discovered. As discussed below, this can be interpreted as meaning that the sensitivities are insufficient. However, it might also indicate that a significant fraction of 100-TeV gamma-ray sources are leptonic and that the accelerators of PeV hadrons are so thin that they produce little gamma-ray and neutrino emission. This suspicion is strengthened by observations that indicate that young and middle-aged pulsars have high efficiencies for producing TeV gamma rays via leptonic processes~\cite{2011ApJ...741...40T, 2014JHEAp...1...31T, 2017Sci...358..911A, 2017PhRvD..96j3016L, 2018A&A...612A...2H, 2019PhRvD.100d3016S, 2021PhRvD.104j3002D, 2021ApJ...911L..27A, 2021JCAP...08..010S, 2022A&A...660A...8B, 2022ApJ...930L...2D, 2022arXiv220500521J, 2022MNRAS.511.1439F, 2022arXiv220711178M}.

Measurements of diffuse gamma-ray and neutrino emission are also important for understanding the hadronic source population. Recently, Tibet AS$\gamma$ made the first observations of diffuse gamma rays near the 100-TeV range~\cite{2021PhRvL.126n1101A}. The flux may be  originating from the ISM~\cite{2018PhRvD..98d3003L, 2022arXiv220315759D}, where CRs lose a small fraction, $f_{\pi, \rm ISM}$, of energy to pions, where
\begin{equation}
    f_{\pi, \rm ISM}\sim 10^{-2}\left(\frac{X_{\rm ISM}}{1~\rm g~cm^{-2}}\right),
    \label{eq:f_pi}
\end{equation}
which is energy-dependent. This is derived using $f_{\pi, \rm ISM}\sim \tau_{\rm ISM}^{\rm esc}/\tau^{pp}_{\rm ISM}$ and $\tau^{pp}_{\rm ISM} = (\kappa_{pp}n_{\rm ISM}^{\rm gas}\sigma_{pp}c)^{-1}$, where $\kappa_{pp}=0.5$ is the inelasticity and $\sigma_{pp}$ is the $pp$ cross section, for which we use 50~mb~\cite{2006PhRvD..74c4018K}. We use $X_{\rm ISM}$ to explicitly note the contribution from the ISM. If we  extrapolate Eq.~(\ref{eq:X-obs}), the grammage for PeV particles is only $\sim0.2$~g~cm$^{-2}$, implying that such CRs lose only a small fraction of their energy, $\sim2\times10^{-3}$, to pion production in the ISM.

Unresolved sources (both leptonic and hadronic) can contribute to the diffuse flux~\cite{2000ApJ...540..923B, 2007APh....27...10P, 2008APh....29...63C, 2014PhRvD..90b3010A, 2016PhRvD..93a3009A, 2018PhRvL.120l1101L, 2021ApJ...919...93F, 2021ApJ...914L...7L, 2022ApJ...928...19V}. The grammage acquired in a source is
\begin{equation}
    X_{\rm src} = 0.1\left(\frac{n_{\rm src}^{\rm gas}}{5~\rm cm^{-3}}\right)\left(\frac{\tau_{\rm src}^{\rm esc}}{\rm 10~kyr}\right)\rm ~g~cm^{-2},
    \label{eq:X_src}
\end{equation}
where $n_{\rm src}^{\rm gas}$ is the gas density at the source and $\tau_{\rm src}^{\rm esc}$ is the escape time (i.e., how long CRs are confined by the source.) The fraction of proton energy lost in sources can be estimated from $X_{\rm src}$, similar to Eq.~(\ref{eq:f_pi}). 

Because source spectra are often harder than the Galactic spectrum, the total source emission might be expected to dominate the diffuse emission at the highest energies. For source emission to be important, $X_{\rm src}$ must be comparable to $X_{\rm ISM}$. Such a possibility has been discussed in the GeV--TeV range, but not been constrained in the PeV range. Below, we show that this is constrained by source counts.

%%%%%%%%%%%%%%%%%%%%%%%%%%%%%%%%%%%%%%%%%%%%%%%%%%%%%%%%%%%%%%%%%%%%%%%%%%%%%%%%
%%%%%%%%%%%%%%%%%%%%%%%%%%%%%%%%%%%%%%%%%%%%%%%%%%%%%%%%%%%%%%%%%%%%%%%%%%%%%%%%

\section{Models of the Hadronic Source Population: Simplified Overview}
\label{sec:analytical-population}

In this section, we introduce a model of the hadronic source population in which the results are \emph{explicitly normalized by the energy-dependent hadronic CR energy budget in  Eq.~(\ref{eq:CR_production})}.  We predict the expected source counts and their gamma-ray and neutrino properties (e.g., luminosities, positions, and fluxes) as well as the total emission from sources. Here we clarify the basic ideas for the $n_{\rm src}^{\rm gas}$--$\tau_{\rm src}^{\rm esc}$ plane (hereafter ``$n$--$\tau$ plane") introduced in this paper.

%%%%%%%%%%%%%%%%%%%%%%%%%%%%%%%%%%%%%%%%%%%%%%%%%%%%%%%%%%%%%%%%%%%%%%%%%%%%%%%%

\subsection{Impulsive Injection}

We first consider the injection of CRs by impulsive events, as expected for SNRs; in the subsequent subsection, we discuss the case where injection is continuous. The total energy produced in CRs per event follows from Eq.~(\ref{eq:CR_production}) and the rate of events that produce them (``CR source rate"), $\Gamma_{\rm CR}$, as 
\begin{equation}
E_{p}^2 \frac{dN_{p}}{dE_{p}}\biggr\rvert_I = \frac{1}{\Gamma_{\rm CR}} E_{p}^2 \frac{d^2N_{p}}{dE_{p} dt}\biggr\rvert_{\rm MW},
\label{eq:cr-proton}
\end{equation}
where $I$ stands for impulsive injection. In the baseline scenario, $\Gamma_{\rm CR}$ is set by the SN rate (0.03 yr$^{-1}$), though it might be that only a subclass of SN explosions produces CRs that reach PeV energies~\cite{2014MNRAS.440.2528M, 2020APh...12302492C}.  We discuss other choices below.

The gamma-ray and neutrino emission from these sources depends on how dense the target is ($n_{\rm src}^{\rm gas}$) and how long CRs are confined ($\tau_{\rm src}^{\rm esc}$).  We assume that the CR energy is impulsively injected, i.e., in a timescale $\Delta t$ shorter than $\tau_{\rm src}^{\rm esc}$ (although Eq.~(\ref{eq:cr-proton}) is a good approximation even if $\Delta t \sim \tau_{\rm src}^{\rm esc}$).  We thus take the confined CR energy to be constant, as described by Eq.~(\ref{eq:cr-proton}), during the time $\tau_{\rm src}^{\rm esc}$ and zero afterwards, when the source CRs mix in with Galactic CRs. The total number of sources is then $\Gamma_{\rm CR}\tau_{\rm src}^{\rm esc}$.

The gamma-ray and neutrino ($\nu+\bar{\nu}$, per flavor) luminosities of a source, normalized to CR data as per Eq.~(\ref{eq:cr-proton}), are 
\begin{equation}
E_\gamma^2 \frac{d^2N_\gamma}{dE_\gamma dt} = \frac{\epsilon_A}{3\tau^{pp}_{\rm src}} E_{p}^2 \frac{dN_{p}}{dE_{p}}\biggr\rvert_{I,\ E_{p} = 10E_\gamma},
\label{eq:gamma-luminosity}
\end{equation}
\begin{equation}
E_\nu^2 \frac{d^2N_{\nu_i}}{dE_\nu dt} = \frac{\epsilon_A}{6\tau^{pp}_{\rm src}} E_{p}^2 \frac{dN_{p}}{dE_{p}}\biggr\rvert_{I,\ E_{p} = 20E_\nu},
\label{eq:nu-luminosity}
\end{equation}
where $\epsilon_A$ takes into account the fact that both the CR and the target material contain heavy nuclei (mostly helium)~\cite{1970Ap&SS...6..377S, 1992ApJ...394..174G, 2009APh....31..341M, 2014ApJ...789..136K}. We use $\epsilon_A=2.6$, as found by Ref.~\cite{2022A&A...659A..57P} for the PeV range, which depends on the uncertain CR composition; see also Ref.~\cite{2022A&A...661A..72B}. We refer to them as luminosities \emph{near} specific energies, which means integrated over bins with $\Delta \ln E \sim 1$. We do not include gamma-ray attenuation; at 100~TeV, the effect is at most a factor of $\sim$2 and certainly smaller for most sources~\cite{2006ApJ...640L.155M, 2016PhRvD..94f3009V, 2017EPJC...77...66C}, but note that this becomes increasingly important at higher energies.

%%%%%%%%%%%%%%%%%%%%%%%%%%%%%%%%%%%%%%
\begin{figure*}
\includegraphics[width=2\columnwidth]{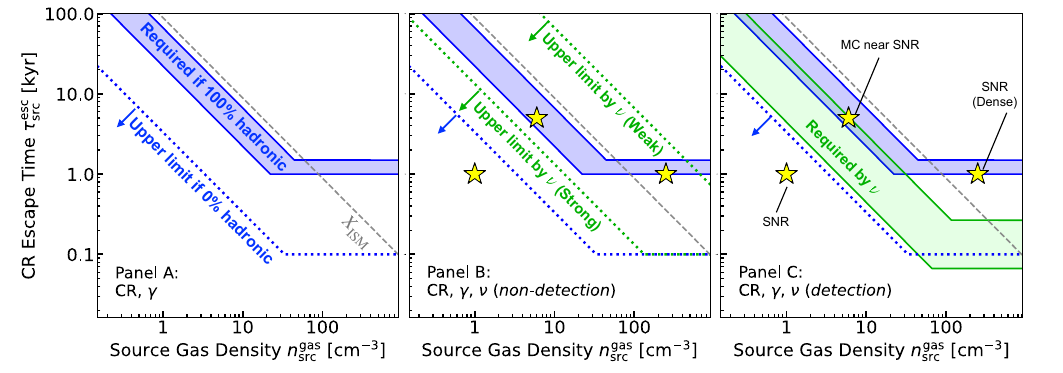}
    \caption{Schematic constraints in the $n$--$\tau$  plane for PeV sources. Panel A considers hypothetical gamma-ray source data in light of real CR data.  If all gamma-ray sources are hadronic, then the sources must lie within the blue band; if none are, then the sources must lie below the blue dotted line.  Regardless of the interpretation of the nature of the gamma-ray sources, the upper right of the plane is excluded. Panels B and C compare hypothetical neutrino source data (green bands/lines, with similar interpretations).  The dashed gray line indicates where the column density encountered by CRs in the source ({$n_{\rm src}^{\rm gas} c\tau_{\rm src}^{\rm esc}$}) equals that encountered in the ISM before escape from the Milky Way ($X_{\rm ISM}$), a relevant scale for comparison.  Stars indicate possible source classes.  For this figure, we assume a CR source rate of $0.03$~yr$^{-1}$. {This new, semi-model-independent plane allows consistent comparisons of CR, gamma-ray, and neutrino data.} See text for details.}
    \label{fig:nt-intro}
\end{figure*} 

%%%%%%%%%%%%%%%%%%%%%%%%%%%%%%%%%%%%%%

Quantitatively, we find the total number of gamma-ray and neutrino sources in the Milky Way to be :
\begin{equation}
\mathcal{N}_\gamma = \mathcal{N}_\nu \sim 300
\left(\frac{\Gamma_{\rm CR}}{0.03~\rm yr^{-1}}\right)
\left(\frac{\tau_{\rm src}^{\rm esc}}{10~\rm kyr}\right).
\label{eq:N_tot}
\end{equation}
These sources have luminosities 
\begin{equation}
\begin{split}
E_\gamma^2 \frac{d^2N_\gamma}{dE_\gamma dt}\biggr\rvert_{E_\gamma = 100~\rm TeV} &\sim 10^{32}~{\rm erg~s^{-1}}
\\
&\times\left(\frac{\Gamma_{\rm CR}}{0.03~\rm yr^{-1}}\right)^{-1}
\left(\frac{n_{\rm src}^{\rm gas}}{\rm cm^{-3}}\right),
\label{eq:simple-gamma}
\end{split}
\end{equation}
\begin{equation}
\begin{split}
E_\nu^2 \frac{d^2N_{\nu_i}}{dE_\nu dt}\biggr\rvert_{E_\nu = 50~\rm TeV}  &\sim 5\times 10^{31}~{\rm erg~s^{-1}}
\\
&\times\left(\frac{\Gamma_{\rm CR}}{0.03~\rm yr^{-1}}\right)^{-1}
\left(\frac{n_{\rm src}^{\rm gas}}{\rm cm^{-3}}\right).
\label{eq:simple-nu}
\end{split}
\end{equation}
The details of the spectrum shape are not important here, as we only focus on a range of energies near $E_p\sim1$~PeV. For comparison, the gamma-ray luminosity of the Crab Nebula at 100~TeV is $\sim 5\times 10^{32}$~{erg~s$^{-1}$}~\cite{2021Sci...373..425L}.

The detection horizon depends on the experimental sensitivity, $F_{\rm lim}$, as
\begin{equation}
d_{\rm lim} = \sqrt{\frac{1}{4\pi F_{\rm lim}}E_i^2 \frac{d^2N_i}{dE_i dt}},
\label{eq:d_lim}
\end{equation}
where $i=\gamma$ or $\nu$. A fraction $\sim \xi_{\rm f}(d_{\rm lim}/R_{\rm gal})^{2}$ of the total source population can be detected, where we assume that we are at the center of a thin disk and the factor $\xi_{\rm f}$ accounts for the limited fraction of the sky (here we take $\xi_{\rm f} \sim 1/3$ ). The expected gamma-ray source count at 100~TeV is then
\begin{equation}
\mathcal{N}_{\rm det}\sim3\left(\frac{\tau_{\rm src}^{\rm esc} \, n_{\rm src}^{\rm gas}}{10~\rm kyr~cm^{-3}}\right)\left(\frac{F_{\rm lim}}{10^{-13}~\rm erg~cm^{-2}~s^{-1}}\right)^{-1},
\label{eq:N-det}
\end{equation}
where we take the Galactic radius to be $R_{\rm gal} =$15~kpc. {(The above approximation is for demonstrative purpose; in the next section we include the source spatial distribution and the full detector sensitivities as functions of source positions.)} 

Figure~\ref{fig:nt-intro} shows the concepts behind using the $n$--$\tau$  plane to constrain hadronic PeVatrons.  Here we illustrate a simple version of the figure with hypothetical data, explaining the implications one by one.  In the next section, we use real data. We start by focusing on $n_{\rm src}^{\rm gas}$ and $\tau_{\rm src}^{\rm esc}$, and then discuss the effects of changing the CR source rate, $\Gamma_{\rm CR}$, and the source size, $R_{\rm src}$.

\begin{itemize}

\item \emph{Cosmic-ray data:} 
By construction, any source class in this plane is consistent with the energy-dependent CR production rate in the PeV range, as in Eq.~(\ref{eq:CR_production}). Hadronic sources that do not reach PeV energies, or those that accelerate PeV particles but do not allow escape (CR hidden sources), are not considered. To derive the CR yield per source, as in Eq.~(\ref{eq:cr-proton}), we assume a CR source rate of $\Gamma_{\rm CR}=0.03$~yr$^{-1}$ (alternatives are discussed below).

\item \emph{Gamma-ray data:}
Consider a hypothetical gamma-ray observation that attains $F_{\rm lim}=10^{-13}~\rm erg~cm^{-2}~s^{-1}$ and observes 10 sources at $E_\gamma=100$~TeV. If all of them are hadronic PeVatrons, Eq.~(\ref{eq:N-det}) indicates that $\tau_{\rm src}^{\rm esc} \, n_{\rm src}^{\rm gas} \sim 30~\rm kyr~cm^{-3}$, with smaller values of $\tau_{\rm src}^{\rm esc} \, n_{\rm src}^{\rm gas}$ allowed if there is a smaller fraction of hadronic sources (and hence a larger fraction of leptonic sources).  In panel A, we show two extreme cases, where 100$\%$ or 0$\%$ of them are hadronic, corresponding to the optimistic and pessimistic cases in Fig.~\ref{fig:schematic}.  Gamma-ray data alone are usually insufficient to decisively determine if a source is hadronic or not.

If all of the observed gamma-ray sources are hadronic, then the parameters $n_{\rm src}^{\rm gas}$ and $\tau_{\rm src}^{\rm esc}$ must be within the shaded area enclosed by the solid band labeled ``Required if 100$\%$ hadronic." This area is defined by $\mathcal{N}_{\rm det}=10$, with the width indicating statistical fluctuations. If none of the observed sources are hadronic, $n_{\rm src}^{\rm gas}$ and $\tau_{\rm src}^{\rm esc}$ must be below the dotted line labeled ``Upper limit if 0$\%$ hadronic," obtained by setting $\mathcal{N}_{\rm det}<1$ (ignoring statistical fluctuations).  One could also display parameter spaces that are consistent with intermediate cases (e.g., half of them being hadronic). 

Regardless of the hadronic fraction of gamma-ray sources, the upper right side of the plane is \emph{always} ruled out, due to predicting more sources than observed.

While parameters within the band predict the same \emph{source counts}, the \emph{luminosities} of gamma-ray sources vary along the band. In the upper left, sources are common but less luminous, making the observable sources nearby.  Towards the middle, sources become rarer but more luminous, making the observable sources more distant.  Eventually, $\tau_{\rm src}^{\rm esc}$ becomes smaller than $10/(\xi_f \Gamma_{\rm CR})$, at which point the gamma-ray source number in the field of view becomes fewer than ten. Then it becomes impossible to attain $\mathcal{N}_{\rm det}=10$ for any $n_{\rm src}^{\rm gas}$ by continuing the diagonal, which causes a flattening of the band.  For the same reason, the dotted lines flatten at $1/(\xi_f \Gamma_{\rm CR})$, corresponding to one source.

\item \emph{Neutrino data:}
Consider two hypothetical neutrino experiments, with sensitivities of $F_{\rm lim}=10^{-11}$ and $2\times10^{-13}~\rm erg~cm^{-2}~s^{-1}$ at $E_\nu=50$~TeV, and suppose they detect no sources. Non-detections define dotted lines, and $n_{\rm src}^{\rm gas}$ and $\tau_{\rm src}^{\rm esc}$ must be below them, similar to the case of zero hadronic sources for the gamma-ray data. Those cases are marked as ``Upper limit by $\nu$ (Weak)" and ``Upper limit by $\nu$ (Strong)" in panel B of Fig.~\ref{fig:nt-intro}.  As with the gamma-ray case, the upper right of the plane is ruled out.

Next, consider a hypothetical neutrino experiment with sensitivity of $10^{-13}~\rm erg~cm^{-2}~s^{-1}$ at $E_\nu=50$~TeV, and suppose that it finds one source, which must then be hadronic, defining a robust allowed parameter space. This case is shown in panel C of Fig.~\ref{fig:nt-intro}, marked as ``Required by $\nu$," where we set the neutrino source count to $\mathcal{N}_{\rm det}=1$. We expect the area (statistical fluctuations) to be larger than the gamma-ray case, due to the smaller source number.  And as with the gamma-ray case, the parameters must be in the band, meaning that the lower left of the plane is also ruled out.

\item \emph{Gamma-ray data vs.\ Neutrino data:}
Comparing gamma-ray and neutrino constraints provides information on the fraction of hadronic sources in the gamma-ray observations. In panel B, the ``Neutrino (Weak)" limit would be consistent with a scenario where 100$\%$ of gamma-ray sources are hadronic, except for the bottom right region, where this is excluded. In contrast, the ``Neutrino (Strong)" limit would exclude all scenarios where 100$\%$ of gamma-ray sources are hadronic.

In panel C, there is a narrow parameter space that is consistent with the scenario where 100$\%$ of gamma-ray sources are hadronic, which is where the two bands overlap. The neutrino data rule out the scenario where 0$\%$ of gamma-ray sources are hadronic, as expected, because the dotted blue line is outside the band allowed by the neutrino source detection. This could be true even if the gamma-ray and neutrino telescopes viewed disjoint regions of the sky.

\item \emph{Data vs.\ Theory:}
With observational constraints on $n_{\rm src}^{\rm gas}$ and $\tau_{\rm src}^{\rm esc}$, we can compare these with expectations from specific models of hadronic PeVatrons. In panels B and C, we show three representative cases with star symbols (``theory points"), each explained below.

If all gamma-ray sources are hadronic, then the theory points should be within the allowed band. If none of the sources are hadronic, then the theory points should be below the blue dotted line. Note that even if a theory point is in the ``100$\%$ hadronic" band, it does not always mean that this source class must explain 100$\%$ of the gamma-ray sources,
due to statistical fluctuations. 

\item \emph{Source vs.\ Diffuse:}
The gray dashed lines show the grammage that CRs accumulate in the ISM, or equivalently the fraction of energy lost, as per Eq.~(\ref{eq:f_pi}), obtained by extrapolating Eq.~(\ref{eq:X-obs}) to 1 PeV.  For points above this line, CRs lose more energy at the source than in the ISM before they escape the Galaxy, i.e., $X_{\rm src} > X_{\rm ISM}$. Above this line, the total Galactic emission in gamma rays and neutrinos would be dominated by sources as opposed to diffuse emission. Our full population models (next section) consistently calculate the resolved source counts and also the contribution of sources to the total Galactic emission of gamma rays and neutrinos.

\end{itemize}

The underlying idea of the $n$--$\tau$  plane is allow a consistent comparison of different results --- CR, gamma-ray, and neutrino data, for both source and diffuse emission --- to each other and to theoretical expectations. 

As noted, we \emph{assume} a CR source rate of $\Gamma_{\rm CR}=0.03$~yr$^{-1}$, and then determine the CR energy per source from the observed CR data. If we adopt different values of $\Gamma_{\rm CR}$, most constraints are unchanged, as the source count, as in Eq.~(\ref{eq:N-det}), does not depend on it. However, the flattening of the band is shifted, proportionally to $(\Gamma_{\rm CR})^{-1}$; we show this case below and in the Appendix.

For convenience, we convert the source number into the distance of the nearest source. Assuming a two-dimensional geometry as above, it is $d_{\rm near} = R_{\rm MW}/\sqrt{\mathcal{N}}$, or
\begin{equation}
d_{\rm near} \sim 0.9~{\rm kpc}
\left(\frac{\Gamma_{\rm CR}}{0.03~\rm yr^{-1}}\right)^{-1/2}
\left(\frac{\tau_{\rm src}^{\rm esc}}{10~\rm kyr}\right)^{-1/2}.
\label{eq:d_near}
\end{equation}
This is to be contrasted with the maximum distance telescopes can see, $d_{\rm lim}$, which we note for each experiment in the next section.

An important effect not yet discussed is the source size, $R_{\rm src}$. In the rest of the paper, we adopt a source size of 10~pc unless otherwise noted.  The sensitivities of gamma-ray and neutrino telescopes are degraded for sources that are more extended than the size of the point-spread function (PSF; $\theta_{\rm PSF}$) by a factor of $\sim\theta_{\rm src}/\theta_{\rm PSF}$~\cite{2009ARA&A..47..523H}, where $\theta_{\rm src}$ is the source angular extension. {(Note that this treatment is only approximate; for more detailed calculations see e.g., Ref.~\cite{2018APh...100...69A}.)} This quantitatively changes the band when the point-source detection horizon in Eq.~(\ref{eq:d_lim}) is smaller than  the distance at which $\theta_{\rm src}=\theta_{\rm PSF}$ (which we denote as $d_{\rm size}$). The detection horizon in this case shrinks to
\begin{equation}
    d_{\rm lim} = \frac{L}{4\pi F_{\rm lim}^{\rm PS}d_{\rm size}},
\end{equation}
where $F_{\rm lim}^{\rm PS}$ is the sensitivity for point sources. The dependence of  horizon distance on luminosities changes to $d_{\rm lim}\propto L$, resulting in $\mathcal{N}_{\rm det} \propto (d_{\rm lim})^2\tau_{\rm src}^{\rm esc}\propto L^2\tau_{\rm src}^{\rm esc} \propto (n_{\rm src}^{\rm gas})^2\tau_{\rm src}^{\rm esc}$, so the source count does not depend linearly on $X_{\rm src}$. 

Having introduced the basics of this figure, the following questions should be addressed:

\begin{itemize}

\item \emph{What sets the range of parameters plotted?}
While both $n_{\rm src}^{\rm gas}$ and $\tau_{\rm src}^{\rm esc}$ can vary over many orders of magnitude, we only show limited parameter ranges.  We encourage new work to sharpen our choices.

There are considerations that set the largest values that the parameters can take.  Sources can have $n_{\rm src}^{\rm gas}$ larger than displayed in Fig.~\ref{fig:nt-intro}. However, we show below that $n_{\rm src}^{\rm gas} \gtrsim 1000$~cm$^{-3}$ is firmly ruled out by  existing IceCube data, except for very short escape time ($\lesssim$0.5~kyr).  For $\tau_{\rm src}^{\rm esc}$, an upper bound comes from a physical argument. Particle diffusion is slowest in the Bohm diffusion regime, where the coefficient is $c R_L/3$, which is $3\times 10^{28}$~cm$^2$~s$^{-1}$ for a 1-PeV particle in a $\mu$G field, where $R_L$ is the Larmor radius. If such a slow diffusion is sustained over 100-pc scale (note that this is extremely optimistic), we obtain $\tau_{\rm src}^{\rm esc} \sim 100$~kyr. Such a large value, although cannot be excluded from first principles, is certainly too high, though we display it in the figure to show present constraints without theoretical bias.

There are also considerations that set the smallest values that the parameters can take. No immediate physical arguments prohibit sources from having small values of $n_{\rm src}^{\rm gas}$, though some matter is required to support the magnetic fields that accelerate (and confine) CRs.  The gas densities can be much smaller than 1~cm$^{-3}$ if, for example, a SN occurs in a cavity region where material has been blown out due to stellar winds or SN explosions.  For $\tau_{\rm src}^{\rm esc}$, the absolute minimum would be the time needed to accelerate particles to energies beyond 1 PeV, which in principle could be as small as $\sim R_L/c$ {in any scenario of particle acceleration~\cite{2002PhRvD..66b3005A}. This is only a few years for a $\mu$G magnetic field and can be even} smaller for a stronger field. These small values are not shown in the figure, but may be physically plausible.

\item \emph{How should the complexity of sources be incorporated into phenomenological descriptions with $n_{\rm src}^{\rm gas}$ and $\tau_{\rm src}^{\rm esc}$?}
The environments of CR sources and the production sites of gamma rays should be quite complicated. In the case of SNRs, gamma rays may be produced by either diffuse matter or gas clumps inside the shell or in their close vicinity, with all of these components likely being highly inhomogeneous~\cite{1994A&A...285..645A, 2000ApJ...538..203B, 2010ApJ...723L.122U, 2010ApJ...708..965Z, 2011JCAP...05..026C, 2012ApJ...744...71I, 2013APh....43...71A, 2019MNRAS.487.3199C, 1934PNAS...20..259B, 2020ApJ...904L..24S, 2021Ap&SS.366...58S, 2021ApJ...919L..16Y, 2022ApJ...925..193Y}. Moreover, the accelerator of particles and the emitter of gamma rays might be physically distinct. For example, an SNR may accelerate protons, which can interact with nearby molecular clouds (MCs), emitting gamma rays and neutrinos~\cite{1996A&A...309..917A, 2007ApJ...665L.131G, 2008A&A...481..401A, 2009MNRAS.396.1629G, 2009ApJ...707L.179F, 2011MNRAS.410.1577O, 2012ApJ...749L..35U, 2021MNRAS.503.3522M}.

For a single source class, we can separately consider multiple ``emitting regions," each characterized by different combinations of $n_{\rm src}^{\rm gas}$ and $\tau_{\rm src}^{\rm esc}$. In the case of SNRs, one can separately consider SNR shells, wind bubbles, wind cavities, and nearby molecular clouds, for example. Due to the likely inhomogeneities of the densities, the parameter $n_{\rm src}^{\rm gas}$ should be regarded as an average gas density that PeV CRs encounter. Formally, it should be calculated by convolving the gas density with the spatial distribution of CRs, which requires a detailed understanding of CR propagation and the gas distribution. More simply, we can separate the source into sub-volumes of density $n_{v}$ and filling fraction $f_v$ and expect $n_{\rm src}^{\rm gas}\sim \sum_v f_v n_{v}$, with $\sum_vf_v=1$. 

\item \emph{Where do realistic models lie in this plane?}
Below, we discuss three concrete cases, each marked in Fig.~\ref{fig:nt-intro}.  Although we focus on SNRs as illustrative examples, our framework can be applied to more general source classes, for which we encourage further work.

\begin{enumerate}

    \item {\bf SNR}: The most plausible candidate for the source of PeV hadrons are SNRs. Indeed, both GeV and TeV data support the scenario where SNRs accelerate protons to TeV scales~\cite{2012A&A...538A..81M, 2013Sci...339..807A,  2016ApJ...816..100J, 2018A&A...612A...3H}, although it remains unknown if the maximum energy can reach the PeV range. If produced, PeV protons are expected to escape in the very early phase of the SNR evolution, $\tau_{\rm src}^{\rm esc}\sim$~1~kyr, comparable to when the Sedov-Taylor phase starts, although many details are uncertain~\cite{2011MNRAS.415.1807D, 2016PhRvD..94h3003D, 2016MNRAS.461.3552N, 2018MNRAS.475.5237G, 2019MNRAS.484.2684N, 2019MNRAS.490.4317C, 2020A&A...634A..59B, 2021A&A...654A.139B, 2021PhRvD.104h3028K, 2022ApJ...924...45S}. Acceleration to the PeV scale might take place on timescales much shorter than kiloyears \cite{2003A&A...403....1P, 2013MNRAS.435.1174S, 2015APh....69....1C, 2018MNRAS.479.4470M, 2021ApJ...922....7I} and $\tau_{\rm src}^{\rm esc}$ includes the time particles are in the vicinity of the accelerator.  This choice of $\tau_{\rm src}^{\rm esc}$ is likely optimistic, given the escape of PeV particles from the shock might be as short as $\sim$10 yrs~\cite{2013MNRAS.435.1174S} and the escape from the larger surroundings are highly uncertain. If PeV protons interact with the average gas densities in the ISM, then $n_{\rm src}^{\rm gas}\sim$~1~cm$^{-3}$.

    \item {\bf SNR (Dense)}: For a handful of shell-type SNRs, TeV gamma rays are spatially coincident with gas clouds, supporting a hadronic origin for the gamma rays: RX 1713.7-3946~\cite{2007ApJ...661..236A, 2012ApJ...746...82F, 2021ApJ...915...84F}, Vela Jr.~\cite{2017ApJ...850...71F}, HESS J1731-347~\cite{2014ApJ...788...94F}, and RCW86~\cite{2019ApJ...876...37S}. They are young ($\simeq1$$-$$5$~kyr) and have high target gas densities of $\gtrsim$10--100~cm$^{-3}$, although the latter significantly depends on the volume filling factor of dense gas, which is usually highly uncertain. Here, we take RX J1713.7-3946 as an example case. Multi-wavelength modeling of this SNR suggests high-density ($2.5\times10^4$~cm$^{-3}$) clumps with a volume filling factor $10^{-2}$, embedded in low-density gas ($\sim 10^{-2}$~cm$^{-3}$), for an average density of $n_{\rm src}^{\rm gas}\sim$~250~cm$^{-3}$ in the 10-pc shell~\cite{2022arXiv220512276F} (see also Ref.~\cite{2019MNRAS.487.3199C} for a detailed numerical study). The escape of PeV particles is model-dependent. Given the lack of $>$~10~TeV gamma-ray emission from this object, $\tau_{\rm src}^{\rm esc}$ is likely smaller than its age (1.4 kyr). {If the escape time coincides with the Sedov time, it would be smaller for SNRs in dense environments (as $\propto (n_{\rm src}^{\rm gas})^{-1/3}$). The actual escape time might be even shorter, as discussed above.} We optimistically take $\tau_{\rm src}^{\rm esc}=1$~kyr , as in the previous case, {but the above uncertainties should be kept in mind.}

    \item {\bf MC near SNR}: Emission may be produced by CRs that escape from the accelerators and  diffuse around them, interacting with a massive gas cloud or clouds. The duration is determined by the local propagation of CRs. As a reference, we consider a molecular cloud with a mass of $M_{\rm cl}=10^5~M_\odot$ and a size of $R_{\rm cl} = 20$~pc at a distance of $d_{\rm cl}=$~50~pc from an SNR. With a diffusion coefficient of $D=10^{29}~$cm$^2$~s$^{-1}$ {at 1~PeV} (ten times smaller than the ISM average), the diffusion time is $\tau_{\rm src}^{\rm esc}\sim(d_{\rm cl})^2/D\sim$~5~kyr. {(Note that the use of an isotropic diffusion coefficient can be a crude approximation close to the source; more work is needed to theoretically evaluate the propagation of PeV particles in the source vicinity.)} The gas density of this MC is very high, $\simeq100~{\rm cm^{-3}}$, but the volume filling fraction of this is $\sim(R_{\rm cl}/d_{\rm cl})^3\sim0.06$, resulting in a modest value of $n_{\rm src}^{\rm gas}\sim$~6~cm$^{-3}$.
    
\end{enumerate}

\item \emph{How should the variation among sources be treated?}
Both $n_{\rm src}^{\rm gas}$ and $\tau_{\rm src}^{\rm esc}$ should vary among sources. In the case of SNRs, the measured gas densities are known to vary by more than an order of magnitude among different sources~\cite{2013A&A...553A..34D, 2019ApJ...874...50Z, 2020PASJ...72...72S}. There are two theoretical possibilities to account for this source-to-source variation. 

First, among all varieties of a source class, only a subclass with specific $n_{\rm src}^{\rm gas}$ and/or $\tau_{\rm src}^{\rm esc}$ values might be able to produce PeV hadronic CRs. In this case, an appropriate value of $\Gamma_{\rm CR}$ should be chosen to take into account the lower rate. A smaller value of $\Gamma_{\rm CR}$ does not move the ``theory points," but does change the allowed bands and limit lines, as discussed above.

Second, sources with a wide range of $n_{\rm src}^{\rm gas}$ and/or $\tau_{\rm src}^{\rm esc}$ values may produce PeV hadronic CRs, and only those with sufficiently large values of these parameters may be detectable in gamma rays and neutrinos. Formally, one should introduce a probability density $d^2\mathcal{P}/dn_{\rm src}^{\rm gas}d\tau_{\rm src}^{\rm esc}$. Simply, we may use the $n_{\rm src}^{\rm gas}$ and $\tau_{\rm src}^{\rm esc}$ values that maximally contribute to the source counts. We can instead imagine a situation where the distributions are approximated as bimodal. Such a scenario can be studied by using Fig.~\ref{fig:nt-intro} as a guide. For example, a fraction $f_{\rm dense}$ of SNR may be in dense regions, like in the ``SNR (Dense)" point, while the remainder are in a normal environment, like in the ``SNR" point.  Figure~\ref{fig:nt-intro} shows that the ``SNR (Dense)" and ``SNR" models predict that $\sim100\%$ and $\sim0\%$ of gamma-ray sources are hadronic, respectively. Then, in total we expect that $\sim100f_{\rm dense}\%$ are hadronic. This consideration can be extended to more general distributions. The key point is that the applicability of Fig.~\ref{fig:nt-intro} is wide, despite the fact that it assumes a universal $n_{\rm src}^{\rm gas}$ and $\tau_{\rm src}^{\rm esc}$.

\end{itemize}

By construction, the $n$--$\tau$ plane aims to constrain the source class that dominates PeV CR energy budget. It is quite plausible that other source classes produce gamma-ray (or even neutrino) fluxes but make subdominant contributions to the hadronic CR flux. For a source class that produce CRs at a rate a factor of $\xi$ smaller than in Eq.~(\ref{eq:CR_production}), all the constrains would approximately be shifted rightwards by a factor of $\xi$.

%%%%%%%%%%%%%%%%%%%%%%%%%%%%%%%%%%%%%%%%%%%%%%%%%%%%%%%%%%%%%%%%%%%%%%%%%%%%%%%%

\subsection{Continuous Injection}

The case of continuous production is straightforwardly obtained from the results above. The luminosity of each source follows from Eq.~(\ref{eq:CR_production}) and the number of objects producing CRs, $\mathcal{N}_{\rm CR}$, as
\begin{equation}
E_{p}^2 \frac{d^2N_{p}}{dE_{p} dt}\biggr\rvert_C = \frac{1}{\mathcal{N}_{\rm CR}} E_{p}^2 \frac{d^2N_{p}}{dE_{p} dt}\biggr\rvert_{\rm MW},
\end{equation}
where $C$ stands for continuous injection. The CR energy spectrum is 
\begin{equation}
E_{p}^2 \frac{dN_{p}}{dE_{p}}\biggr\rvert_C = \frac{\tau_{\rm src}^{\rm esc}}{\mathcal{N}_{\rm CR}} E_{p}^2 \frac{d^2N_{p}}{dE_{p} dt}\biggr\rvert_{\rm MW}.
\end{equation}
Here, the sources are assumed to be in steady-state conditions, which necessitates the {escape} time $\tau_{\rm src}^{\rm esc}$ to be less than the age of the source (more than Myr-scales for the case of star clusters). Then Eqs.~(\ref{eq:gamma-luminosity}) and (\ref{eq:nu-luminosity}) can be used to calculate luminosities, with $I$ replaced with $C$. 

In the continuous case, the total Galactic source number $\mathcal{N}$ is fixed, and larger values of $n_{\rm src}^{\rm gas}$ and $\tau_{\rm src}^{\rm esc}$ both lead to larger luminosities, as $L_\gamma \propto n_{\rm src}^{\rm gas}\tau_{\rm src}^{\rm esc}$. In the plane of $n_{\rm src}^{\rm gas}$--$\tau_{\rm src}^{\rm esc}$, the upper right region corresponds to more luminous sources. Because the detection area increases with $d_{\rm lim}^2 \propto L_\gamma$, the detectable source count increases accordingly as $\mathcal{N}_{\rm det}\propto \mathcal{N}L_\gamma \propto n_{\rm src}^{\rm gas}\tau_{\rm src}^{\rm esc}$. This is in contrast to the impulsive case, where the total Galactic source number is $\mathcal{N}\propto \tau_{\rm src}^{\rm esc}$ and their luminosities are $L_\gamma\propto n_{\rm src}^{\rm gas}$, although this case also results in $\mathcal{N}_{\rm det}\propto \mathcal{N}L_\gamma \propto n_{\rm src}^{\rm gas}\tau_{\rm src}^{\rm esc}$. Additionally, the band does not flatten in the continuous case.

An example case is stellar winds of massive stars in star clusters~\cite{2020SSRv..216...42B}. Gamma-ray observations support the idea that this source class is continuously injecting CRs into the ISM~\cite{2019NatAs...3..561A}. The extent of the gamma-ray emitting region is often greater than the size of the cluster itself and can exceed $\sim$50~pc. The diffusion in this region is slow. Assuming $D = 10^{29}$~cm$^2$~s$^{-1}$ (ten times smaller than the ISM average), $\tau_{\rm src}^{\rm esc}\sim 5$~kyr. The gas densities are suggested to be high $n_{\rm src}^{\rm gas} \sim 10~$cm$^{-3}$~\cite{2018A&A...611A..77Y,  2020A&A...639A..80S, 2021NatAs...5..465A}.  These points suggest that star clusters are promising PeVatron candidates that can be probed by gamma rays and neutrinos.  (In compact star clusters or more loose young star associations, multiple SNRs can also be produced, which might efficiently accelerate hadrons compared to isolated SNRs~\cite{2005ApJ...628..738H, 2020MNRAS.493.3159G, 2022MNRAS.510.5579B, 2022MNRAS.512.1275V, 2022arXiv220701432V}, which should be viewed as ``impulsive" injection.)

%%%%%%%%%%%%%%%%%%%%%%%%%%%%%%%%%%%%%%%%%%%%%%%%%%%%%%%%%%%%%%%%%%%%%%%%%%%%%%%%
%%%%%%%%%%%%%%%%%%%%%%%%%%%%%%%%%%%%%%%%%%%%%%%%%%%%%%%%%%%%%%%%%%%%%%%%%%%%%%%%

\section{Models of the Hadronic Source Population: Technical Methods}
\label{sec:model-population}

In this section, we describe the methods for our Monte-Carlo simulations of the population of hadronic gamma-ray and neutrino sources, with the results and interpretation given in the next section.  We continue to use the phenomenological parameters $\tau_{\rm src}^{\rm esc}$ and $n_{\rm src}^{\rm gas}$ as in Eqs.~(\ref{eq:N_tot}), (\ref{eq:simple-gamma}), and (\ref{eq:simple-nu}).

For each parameter set $\{\tau_{\rm src}^{\rm esc}, n_{\rm src}^{\rm gas}, R_{\rm src}\}$, we run the simulation $10^4$ times, each time sampling $\Gamma_{\rm CR}\tau_{\rm src}^{\rm esc}$ sources and calculating the distributions of observables. We place sources randomly in the Milky Way plane, following the radial distribution of initial positions of neutron stars used in Ref.~\cite{2010A&A...510A..23S} (the ``F06 model"~\cite{2006ApJ...643..332F}), and assuming a Gaussian distribution for the height above the plane, with a standard deviation of 30 pc~\cite{2006ApJ...643..332F}. We then calculate the source luminosities with Eqs.~(\ref{eq:simple-gamma}) and (\ref{eq:simple-nu}).  We include a variety of corrections, including how the search sensitivities depend on declination and the extent of the sources, as discussed above.  We set the maximum detectable angular size to be $\theta_{\rm max}=5^\circ$, which excludes sources closer than 100~$(R_{\rm src}/10~\rm pc)$~pc.

In the next section, we compare several quantities with present data, while here we only describe what we calculate and how.  For predictions to agree with observations, we require that the observed values be between the 10th and 90th percentiles.  To disagree, we require that the observed value be less than the 10th percentile. This is conservative because we consider a parameter set to be ruled out only if more than 90$\%$ of the predicted range disagrees with observations. 

\begin{itemize}

\item \emph{Source counts from LHAASO at 100~TeV:}
LHAASO reports the detection of 12 Galactic sources~\cite{2021Natur.594...33C}. This search covers a sky region of $-15^\circ<\delta<75^\circ$. Because the details of the sensitivities are not reported, we assume a sensitivity of 0.4 Crab at 100~TeV at best ($4\times10^{-13}$~erg~cm$^{-2}$~s$^{-1}$), which is comparable to the minimal flux reported in Ref.~\cite{2021Natur.594...33C}. We assume a PSF size of $\theta_{\rm PSF}=0.3^\circ$~\cite{2021ChPhC..45b5002A}.  The sensitivities are degraded closer to the edge of observable sky region; due to the lack of detailed information, we assume a declination dependence of the sensitivity as in the 2HWC survey for a $E^{-2.5}$ spectrum~\cite{2017ApJ...843...40A} . This flux sensitivity corresponds to a distance range of 1.5~kpc for $\Gamma_{\rm CR}=0.03$~yr$^{-1}$ and $n_{\rm src}^{\rm gas}=1$~cm$^{-3}$, as per Eq.~(\ref{eq:simple-gamma}).  This distance being moderately small indicates that LHAASO and future experiments have many more sources to discover.

\item \emph{Source counts from IceCube:}
The latest results from full-sky point-source searches are reported in Ref.~\cite{2020PhRvL.124e1103A}, which uses ten years of track-like events data and looks for the clustering of neutrino events over background. Additional searches with seven years of cascade events data are presented in Ref.~\cite{2019ApJ...886...12A}, which is more sensitive to the sources located in the southern sky. Both searches reported the non-detection of any Galactic sources. {We use these two papers to constrain models. They} report ``5$\sigma$ discovery potential" flux levels as a function of source declination, for two assumed spectrum slopes, $-2.0$ and $-3.0$. We use the former to be conservative, as we are fixing the flux near 50 TeV. {(Using the latter would make the IceCube constraints better by a factor of $\sim2$.)} We consider sources to be detected when the source flux at 50~TeV exceeds 5$\sigma$ discovery potential of either track or cascade searches; tracks are more sensitive for a large portion of the sky, while cascade are slightly better for sources at $\sin\delta \lesssim -0.3$ {as the detector is at the South Pole}. The best detectable flux level reaches $10^{-12}$~erg~cm$^{-2}$~s$^{-1}$ at 50~TeV~\cite{2020PhRvL.124e1103A}. {Had we instead used the ``differential sensitivity", the detectable flux need to be larger by a factor of $\sim3$, and IceCube constraints gets worse by the same factor. It would, however, underestimate the power of IceCube, because lower-energies emission should also contribute to the test statistics for detection.} The IceCube analysis method integrates the significance over energy, putting more weight on high energies~\cite{2019ApJ...886...12A, 2020PhRvL.124e1103A}.  Based on our estimates, the most important energy range is somewhat below the 50 TeV we use, but the sensitivity calculations can be improved in future work.  We assume {$\theta_{\rm PSF}=0.5^\circ$} for track and $\theta_{\rm PSF}=10^\circ$ for cascade events. This flux sensitivity corresponds to a distance horizon of 0.6~kpc for $\Gamma_{\rm CR}=0.03$~yr$^{-1}$ and $n_{\rm src}^{\rm gas}=1$~cm$^{-3}$ (Eq.~\ref{eq:simple-nu}). This distance being very small indicates that IceCube has not yet significantly probed Milky Way sources.

\end{itemize}

In addition to source counts, we calculate the total emission from the Milky Way plane:

\begin{itemize}

\item \emph{Total gamma-ray flux from sources:}
Tibet~AS$\gamma$ reported the detection of emission from the Milky Way plane~\cite{2021PhRvL.126n1101A}. We use the measured diffuse intensity, $5\times10^{-8}$~GeV~cm$^{-2}$~s$^{-1}$~sr$^{-1}$ at $E_\gamma=$100~TeV for the region of $25^\circ < l <100^\circ$ and $|b|<5^\circ$, which includes the truly diffuse ISM emission and contributions from \emph{unresolved} sources, but excludes the contribution from \emph{resolved} (known) TeV sources. The total emission from known sources is on the order of 10$\%$ level~\cite{2021PhRvL.126n1101A, 2018PhRvD..98d3003L}, so we do not correct for this. We calculate the sum of 100-TeV gamma-ray fluxes from simulated sources in the region observed by Tibet, and require that it does not exceed the measured flux. 

\item \emph{Total neutrino flux from sources:}
IceCube {and ANTARES have} searched for the diffuse emission of neutrinos from the Milky Way plane~\cite{2018ApJ...868L..20A, 2019ApJ...886...12A}, which is the sum of the truly diffuse flux plus the contributions from sources. Unlike Tibet, {they have} no sources to be removed. {We use the} upper limit at $E_\nu=50$~TeV ({$1.5\times10^{-7}$}~GeV~cm$^{-2}$~s$^{-1}$, $\nu+\bar{\nu}$, all flavor)~\cite{2018ApJ...868L..20A} obtained using templates for diffuse neutrino emission from Refs.~\cite{2015ApJ...815L..25G, 2017PhRvL.119c1101G}. Note that the latest cascade search reported a 2$\sigma$ level detection, with the best-fit flux about half of this upper limit~\cite{2019ApJ...886...12A}. We calculate the sum of 50-TeV neutrino fluxes from the Milky Way plane ($|b|<5^\circ$) sources, and require that it does not exceed the reported upper limit.

\end{itemize}

By using the observations noted above, we connect the CR energy budget near $E_p=$ 1~PeV to gamma-ray data near $E_\gamma=$~100~TeV and neutrino data near $E_\nu=$~50 TeV, focusing on narrow energy bins that are connected by typical kinematic relations.  To extend our discussions to somewhat lower energies, we make an additional calculation:

\begin{itemize}

\item \emph{Source counts from HAWC at $E_\gamma=7$~TeV:}
We calculate the expected source counts for the latest survey, 3HWC~\cite{2020ApJ...905...76A}, which contains 63 sources (removing two sources that are known to be extragalactic). We calculate the source flux by extrapolating Eq.~(\ref{eq:simple-gamma}) with $d^2N_\gamma/(dE_\gamma dt) \propto (E_\gamma)^{-2.37}$, motivated by the CR production rate, although the caveats discussed below Eq.~(\ref{eq:CR_production}) should be kept in mind. We adopt the differential sensitivity near 7~TeV quoted in Ref.~\cite{2020ApJ...905...76A} for a spectrum index of $-2.5$ and assume $\theta_{\rm PSF}=0.3^\circ$. The best sensitivity reaches 3~$\times10^{-13}$~erg~cm$^{-2}$~s$^{-1}$ (at 7~TeV).

\end{itemize}

We also calculate expectations for future data.

\begin{itemize}

\item \emph{Expected source counts for IceCube-Gen2:}
The sensitivity depends on the details of the detector and search strategies, which are uncertain, but an overall factor of five improvement is expected~\cite{2021JPhG...48f0501A}. We estimate the sensitivity of Gen2 by scaling the 5$\sigma$ discovery-level fluxes from the existing IceCube searches by this factor. {We use $\theta_{\rm PSF}=0.3^\circ$ for track at Gen-2. For cascade, we use the same $\theta_{\rm PSF}$ as for IceCube, although some improvements are expected.}. The detection horizon is 1.3~kpc~$(\Gamma_{\rm CR}/0.03$~yr$^{-1})^{-1/2}$ $(n_{\rm src}^{\rm gas}/1$~cm$^{-3})^{1/2}$.

\item \emph{Expected source counts for other neutrino experiments:}
Several neutrino telescopes, all in the northern hemisphere, are planning to use water instead of ice. Those include KM3NeT, P-ONE, Baikal-GVD, and TRIDENT~\cite{2016JPhG...43h4001A, 2018arXiv180810353B, 2020NatAs...4..913A, 2022arXiv220704519Y}. Those experiments are expected to have better angular resolution than IceCube, improving the power for finding sources, and also covering a large portion of sky outside IceCube's best range.  Although the details are still uncertain, analyses {for KM3NeT/ARCA} suggests a source flux sensitivity of $\simeq$ (0.8--1.6) $\times 10^{-12}$~erg~cm$^{-2}$~s$^{-1}$ {in 6 years}, comparable to IceCube, but over a different range of sky ($-1 < \sin \delta < 0.8$)~\cite{2019APh...111..100A} {(see also Ref.~\cite{2018APh...100...69A})}.  We calculate the source-detection expectations for these combined experiments by assuming a uniform sensitivity over the entire sky (combining with IceCube or Gen2), using two values of 1.2 (``KM3") and 0.16 (``KM3$\times5$") in units of $10^{-12}$~erg~cm$^{-2}$~s$^{-1}$ per flavor at 50~TeV. As a reference, the neutrino sources detectable with ``KM3$\times5$" sensitivity should have a hadronic gamma-ray flux of 0.3~Crab at 100 TeV. The detection horizon for the ``KM3$\times5$" case is still small; 1.5~kpc for $\Gamma_{\rm CR}=0.03$~yr$^{-1}$ and $n_{\rm src}^{\rm gas}=1$~cm$^{-3}$. In all cases, we assume $\theta_{\rm PSF} = 0.3^\circ$, which is conservative for KM3NeT~\cite{2016JPhG...43h4001A}.

\end{itemize}

While gamma-ray telescopes typically have better flux sensitivity, neutrino telescopes have advantages beyond being able to decisively indicated hadronic sources.  First, they cover larger regions of the sky.  Second, sources are typically small compared to the angular resolution, so the flux sensitivity is less subject to being degraded due to the source extent {(see, e.g., Ref.~\cite{2018APh...100...69A})}.

%%%%%%%%%%%%%%%%%%%%%%%%%%%%%%%%%%%%%%%%%%%%%%%%%%%%%%%%%%%%%%%%%%%%%%%%%%%%%%%%
%%%%%%%%%%%%%%%%%%%%%%%%%%%%%%%%%%%%%%%%%%%%%%%%%%%%%%%%%%%%%%%%%%%%%%%%%%%%%%%%

\section{Present Constraints on the Hadronic Source Population}
\label{sec:result-population}

%%%%%%%%%%%%%%%%%%%%%%%%%%%%%%%%%%%%%%
\begin{figure*}
\includegraphics[width=2\columnwidth]{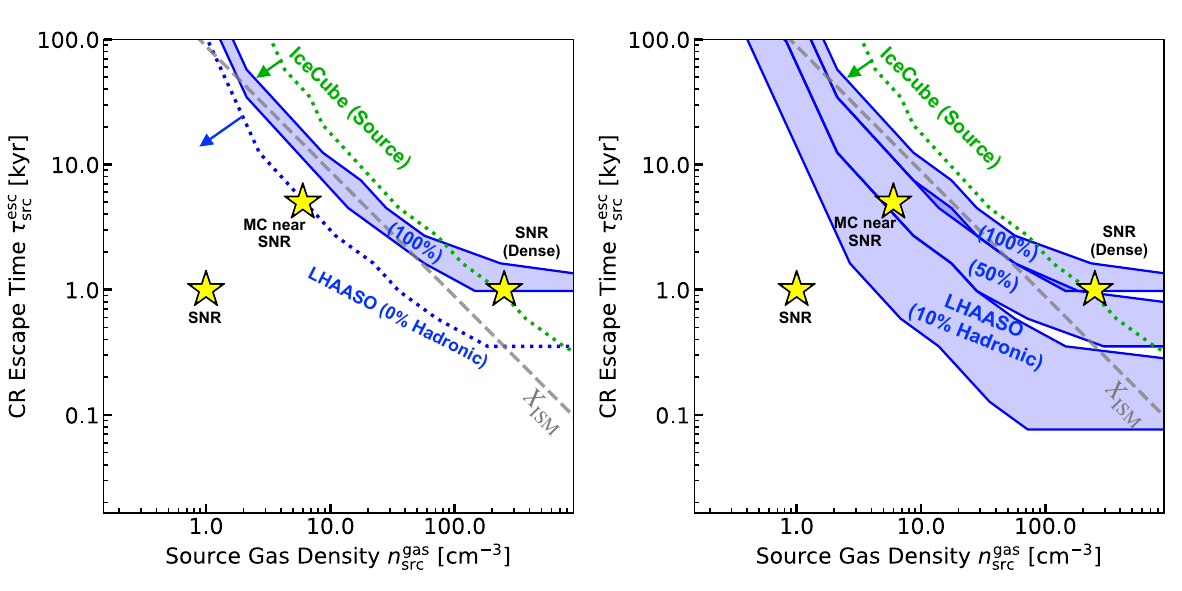}
    \caption{Constraints on PeVatron models from point-source observations, with the models calibrated to hadronic CR data in the PeV range.  Gamma-ray constraints are based on LHAASO source counts (blue band or dotted line) and neutrino constraints are based on IceCube non-detections of sources (green dotted line).  Other aspects follow Fig.~\ref{fig:nt-intro}.  {\bf Left Panel:} Cases where 0$\%$ or 100$\%$ of LHAASO sources are hadronic.  {\bf Right Panel:} Same, but for cases with 10$\%$, 50$\%$, or 100$\%$. {\bf \emph{Key takeaways:}} \emph{(Left) A wide range of models remain viable, with present IceCube constraints allowing between 0\% and 100\% of the LHAASO sources to be hadronic accelerators, although the latter requires a large $n_{\rm src}^{\rm gas}\tau_{\rm src}^{\rm esc}$, so that the source grammage dominates over that of the ISM. (Right) LHAASO is beginning to probe scenarios where SNRs are in high density environments or have a nearby molecular cloud, but not yet the ordinary SNR case, despite our being optimistic on $\tau_{\rm src}^{\rm esc}$.}}
    \label{fig:nt-current}
\end{figure*}
%%%%%%%%%%%%%%%%%%%%%%%%%%%%%%%%%%%%%%

%%%%%%%%%%%%%%%%%%%%%%%%%%%%%%%%%%%%%%
\begin{figure}
    \centering
    \includegraphics[width=\columnwidth]{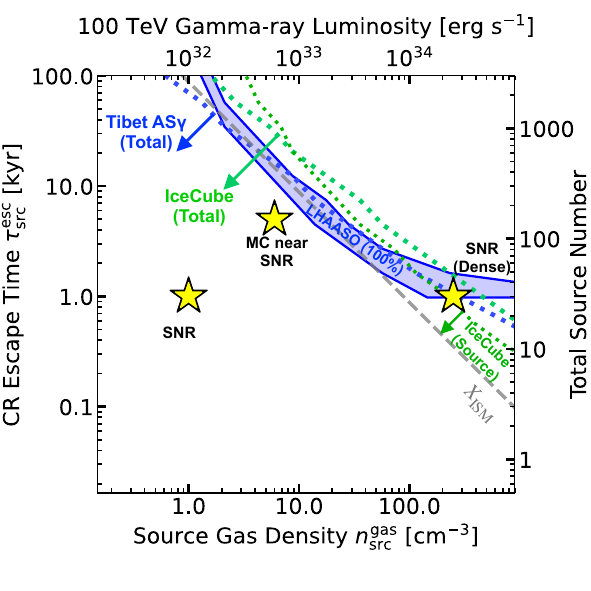}
    \caption{Same as Fig.~\ref{fig:nt-current} (left panel), adding constraints on sources from the Tibet AS$\gamma$ (a measurement) and IceCube (a limit) observations of the Milky Way plane.  The alternate axes also apply to the other main text figures. {\bf\emph{Key takeaway:}} \emph{Constraints from the Milky Way plane emission are important.}}
    \label{fig:nt-current-diffuse}
\end{figure}
%%%%%%%%%%%%%%%%%%%%%%%%%%%%%%%%%%%%%%

%%%%%%%%%%%%%%%%%%%%%%%%%%%%%%%%%%%%%%
\begin{figure}
    \centering
    \includegraphics[width=\columnwidth]{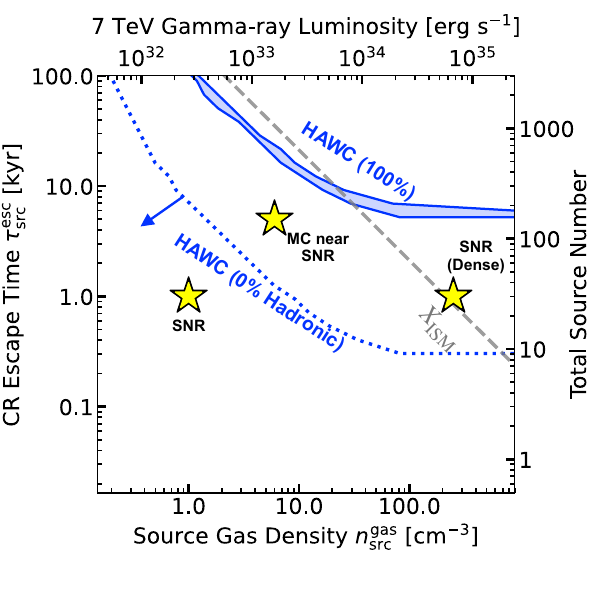}
    \caption{Constraints from HAWC source measurements (at $E_\gamma=7$~TeV). $X_{\rm ISM}$ is scaled to the value for $E_p=70$~TeV.
    {\bf \emph{Key takeaway:}} \emph{Large values of $n_{\rm src}^{\rm gas}$ and $\tau_{\rm src}^{\rm esc}$ are needed to have all HAWC sources to be hadronic, suggesting that leptonic sources may be dominant.}}
    \label{fig:nt-current-hawc}
\end{figure}
%%%%%%%%%%%%%%%%%%%%%%%%%%%%%%%%%%%%%%

In this section, we develop new constraints on hadronic sources using present data on CRs, gamma rays, and neutrinos, as discussed above.  We stress that the $n$--$\tau$  plane is constructed from the energy budget of PeV protons, and is thus consistent with CR observations. 

In short, the LHAASO highest-energy sources are consistent with being either all hadronic or all leptonic. The IceCube non-detections still allow a wide range of PeVatron models, indicating the need for upgraded neutrino telescopes. The source grammage in the PeV range is likely subdominant compared to that in the ISM.  Next, we quantify these points.

Figure~\ref{fig:nt-current} (left panel) shows important results 
on gamma-ray point-source observations. The LHAASO sources may all be hadronic.  We show by a star one theoretical model that is nominally consistent with this, but we note that these models are optimistic and in conflict with IceCube data (below).  It is also possible that none of the LHAASO sources are hadronic.  This is hinted at by the disagreement of the LHAASO band with the fiducial hadronic model of PeV protons from SNRs interacting with gas of density 1~cm$^{-3}$, characteristics of the ISM.  It may thus be that hadronic CR accelerators are so thin (i.e., have small  $n_{\rm src}^{\rm gas}\tau_{\rm src}^{\rm esc}$) that they are not observable multi-messenger sources.  

Figure~\ref{fig:nt-current} (left panel) shows important results 
on neutrino point-source observations.  The IceCube non-detections exclude scenarios with very high average gas densities (and part of the LHAASO ``100$\%$" band on the right), and that is quite useful.  However, many other scenarios remain allowed, including those where all the LHAASO sources are hadronic.  More sensitive neutrino observations are needed.

Figure~\ref{fig:nt-current} (right panel) shows further details about the gamma-ray data.  We show bands assuming that 100\%, 50\%, and 10\% of the LHAASO sources are hadronic PeVatrons. As the total source count is 12, the case of 10$\%$ indicates that about one of them is hadronic. If one source is hadronic, parameters within the ``10$\%$" band are required.  If zero sources are hadronic, the limit is invoked; it is higher than the ``10$\%$" band because we conservatively allow statistical fluctuations.  The ``normal" SNR model is slightly outside the ``10$\%$" band.  If the gas density surrounding typical SNRs is slightly larger than 1~cm$^{-3}$ due to, e.g., compression by the shock, there is a chance that the LHAASO sources include one hadronic accelerator, though the uncertainties are large. {It is also interesting to note that $\tau_{\rm src}^{\rm esc}$ for ``SNR (Dense)" model could be a factor of $\sim(250)^{1/3}\sim6$ smaller than the value in the plot, if it scales with the gas density like the Sedov time; such a scenario would predict $\sim1$ detection by LHAASO.} Overall, this figure suggests that the highest-energy gamma-ray sources are likely dominated by leptonic accelerators, although they might contain one (or even a few) hadronic accelerators.

SNRs have long been the leading candidate for hadronic PeVatrons. However, no SNRs show  evidence of gamma-ray emission beyond 100~TeV, increasing the impression that they (or at least the majority of them) might not accelerate PeV hadrons.  But our results show that even in the scenario where all ordinary SNRs are hadronic PeVatrons, LHAASO most likely does not expect to see 100~TeV gamma rays from them.  LHAASO is only starting to probe scenarios where SNRs have high densities or nearby molecular clouds. {This is consistent with the observational fact that many LHAASO sources are not associated with SNRs, although further observations are needed for in-depth studies of source association because (1) follow-up observations of LHAASO objects could reveal the existence of SNRs and (2) some SNRs observed in TeV gamma rays have no multi-wavelength counterpart.}

Figure~\ref{fig:nt-current} also shows the relative importance of unresolved hadronic sources and truly diffuse emission.  The diagonal dashed line represents $X=X_{\rm ISM}$, where source column densities for CRs are comparable to what they will encounter in the ISM before escaping the Galaxy.  The grammage accumulated in the source ($X_{\rm src}$) could be comparable to $X_{\rm ISM}$; in that case, the LHAASO source counts must be dominated by hadronic sources (except for the case of very high  $n_{\rm src}^{\rm gas}$ and small $\tau_{\rm src}^{\rm esc}$.)  If only a fraction of LHAASO sources are hadronic, $X_{\rm src}$ should be smaller. This means that the total (source + diffuse) Milky Way {luminosity} in the PeV range is likely dominated by the diffuse emission from the ISM.  This conclusion is affected by the uncertain source size (discussed below) and escape time from the Galaxy (a shorter escape time would increase the relative importance of sources). {Note that the above discussion on grammage applies only to the PeV range, which is challenging to observationally probe (contrary to the GeV range, where we have various secondary data). Interestingly, the possibility of $X_{\rm ISM}<X_{\rm src}$ in the PeV range is allowed for some parameter space; exploring this further might lead to new constraints.}

Figure~\ref{fig:nt-current-diffuse} shows the effects of taking into account the total gamma-ray and neutrino fluxes from the Milky Way plane.  The gamma-ray constraint from Tibet AS$\gamma$ is compatible with the constraints by LHAASO source counts.  If one can isolate contributions from the truly diffuse ISM emission (which may be dominant), the constraints will get even stronger.  The constraint from the IceCube neutrino total flux is comparable to the constraint from IceCube source counts; both are in general somewhat weaker than gamma-ray constraints.  Figures~\ref{fig:nt-current-diffuse} also displays the gamma-ray luminosity (the corresponding neutrino luminosity is a factor of 2 smaller than this) and total source number in the Milky Way as a reference; the same scale on Fig.~\ref{fig:nt-current-diffuse} applies to all other figures in the main text.

The source-count constraints get worse if the sources is larger or the observational angular resolution is worse, while the total-emission constraints do not. These indicate the potential power of the diffuse measurements to probe the origin of PeVatrons, especially if they are extended or have low gas densities.  If IceCube confirms the emission from the Milky Way plane, it would define an allowed band in the $n$--$\tau$ plane, similar to that for the LHAASO sources. For the cases considered here, while source crowding could be an issue for cases with a large escape time, the background induced by faint sources in a given direction is never a problem.

Figure~\ref{fig:nt-current-hawc} extends our gamma-ray point-source considerations to somewhat lower energies, where there are more detections. Comparing this to Fig.~\ref{fig:nt-current} shows that there is no overlap between ``100$\%$ Hadronic" areas for LHAASO and HAWC. If the escape times for these two particle energies were the same, having all of the HAWC sources be hadronic would be in conflict with the LHAASO observations. However, HAWC data are based on gamma-ray measurements at 7~TeV, probing $E_p\sim$~70~TeV, while LHAASO data are $E_p\sim$~1~PeV. It is reasonable to expect that the escape time depends on energy, being longer for lower-energy particles. Still, this analysis shows that the required $n_{\rm src}^{\rm gas}$ and $\tau_{\rm src}^{\rm esc}$ values required to explain all the HAWC sources with hadronic accelerators are high, suggesting a significant contribution from leptonic sources or hadronic non-PeVatrons. {Our results are consistent with a more narrowly focused population study by Cristofari \emph{et al.}~\cite{2018MNRAS.479.3415C}, which focused on a standard SNR scenario and found that only limited number of sources should be detected by multi-TeV gamma-ray survey. }

In the Appendix, we explore the following three model variations and their corresponding $n$--$\tau$ planes.

\begin{enumerate}

    \item It may be that only a fraction of supernova explosions produce PeV hadrons. In the scenario where $\Gamma_{\rm CR} = 0.003$~yr$^{-1}$ (10\% of the SN rate), even the ``SNR (Dense)" point predicts that only one of the LHAASO sources are hadronic. This scenario is also allowed by existing IceCube data. 
    
    This scenario should be distinguished from the case where all SNRs are PeVatrons, but only a fraction reside in dense environments. Such a scenario can be studied by Fig.~\ref{fig:nt-current}. For example, it might be that 90\% of SNRs are in a ``normal" environment, while 10\% are in a dense environment. These cases predict 10\% and 100\% gamma-ray source count (Fig.~\ref{fig:nt-current}). In total, hadronic PeVatrons may be $90\%\times0.1+10\%\times1\simeq20\%$ of the highest-energy gamma-ray sources, i.e., 2 or 3 of them.
    
    \item If the emitting regions are larger than 10~pc, more sources escape detection, reducing the capability of source searches. In scenarios where $R_{\rm src} =30$~pc, the standard ``SNR" points are now further outside the ``10$\%$" band, indicating that they do not appear in the highest-energy gamma-ray source count even with a factor of $\sim$10 increase in the gas densities. An important new result is that the possibility that $X_{\rm src} > X_{\rm ISM}$ is allowed over a wider parameter space than the case of 10-pc sources, which suggests that sources could make a dominant contribution to the total Galactic emission.
    
    \item Finally, hadronic accelerators might produce CRs over a long time, rather than impulsively. A scenario that attracts increasing attention is stellar winds in young star clusters over Myr timescales. We consider the scenario where they are hadronic PeVatrons. The total number of OB stars in the Milky Way is estimated to be ${\sim10^5}$~\cite{2018JKAS...51...37S}. The number of OB stars contained in a cluster, $N_{\rm OB}$, is distributed as ${dn/d\log{N_{\rm OB}} \propto 1/N_{\rm OB}}$. This may suggest that clusters with smaller $N_{\rm OB}$ contribute more to the total energetics. However, it is reasonable to think that a larger clusters might have collective effects that make particle acceleration efficient. Here, we consider clusters of $N_{\rm OB}\gtrsim 10^2$ to be efficient PeVatrons~\cite{2018MNRAS.479.5220G}; the number of such clusters are about $10^2$. The kinetic power is $\sim 10^{38}$~erg~s$^{-1}$ per cluster and $\sim 10^{40}$~erg~s$^{-1}$ in total for the Galaxy. Comparison of this to the PeV CR energy budget indicates that 1\% of the kinetic energy must be converted to PeV CRs. This requires 10\% of the energy transferred to CRs with a hard spectrum, such as $E^{-2}$. 
    
    Our estimates for star clusters are encouraging.  They may contribute half of the LHAASO source count, and future Gen2 observations might find as many as three of them. Indeed, one of the LHAASO sources is from the region that contains the Cygnus OB association of massive stars. At the same time, if many of LHAASO sources are associated with star clusters, we might expect more associations than one. This might suggest that we are somewhat in the downward fluctuations. Alternatively, it could be that instead of continuous injection, impulsive events (e.g., SNR shocks embedded in compact clusters) could be the PeV hadron sources. Solid conclusions require more work on both theory and observation.

\end{enumerate}

%%%%%%%%%%%%%%%%%%%%%%%%%%%%%%%%%%%%%%%%%%%%%%%%%%%%%%%%%%%%%%%%%%%%%%%%%%%%%%%%
%%%%%%%%%%%%%%%%%%%%%%%%%%%%%%%%%%%%%%%%%%%%%%%%%%%%%%%%%%%%%%%%%%%%%%%%%%%%%%%%

\section{Future Prospects for Probing the Hadronic Source Population}
\label{sec:result-population-future}

%%%%%%%%%%%%%%%%%%%%%%%%%%%%%%%%%%%%%%
\begin{figure*}
\includegraphics[width=2\columnwidth]{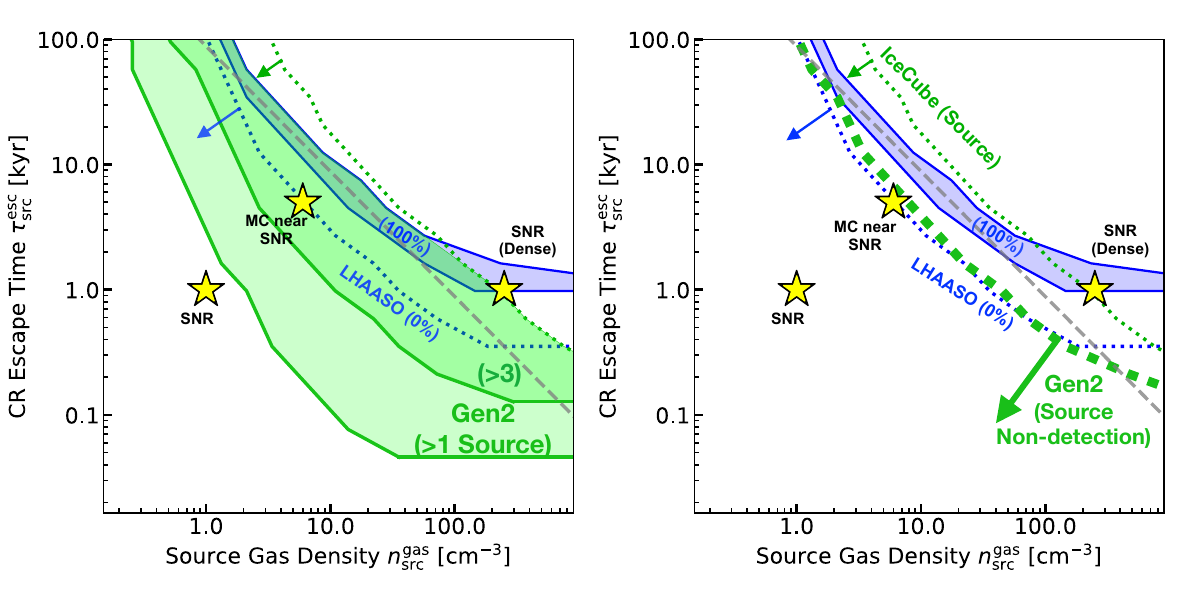}
    \caption{Sensitivity of Gen2 to Galactic hadronic PeVatrons. {\bf Left:} The case of detections.  {\bf Right:} The case of non-detections.  The other features follow Fig.~\ref{fig:nt-current} (left panel).
    {\bf \emph{Key takeaways:}} \emph{Over a wide parameter space, it is promising that Gen2 will find more than one PeVatron, though it may find zero.  Non-detection by Gen2 would rule our models where 100\% of the LHAASO sources are hadronic and also where $X_{\rm src}>X_{\rm ISM}$, but would still allow a wide range of hadronic models.}}
    \label{fig:nt-gen2}
\end{figure*}
%%%%%%%%%%%%%%%%%%%%%%%%%%%%%%%%%%%%%%

In this section, we outline the power of upcoming observations to find hadronic PeVatrons. In Sec.~\ref{subsec:neutrino}, we calculate the prospects for future neutrino observations, which will be the most decisive.  In Sec.~\ref{subsec:gamma_etc}, we discuss improved gamma-ray and other observations, which will be available sooner.  In Sec.~\ref{subsec:population}, we discuss how our population approach, combined with improvements in multi-messenger studies of individual sources, will be critical to solving the long-standing mystery of the Milky Way's hadronic PeVatrons.

%%%%%%%%%%%%%%%%%%%%%%%%%%%%%%%%%%%%%%%%%%%%%%%%%%%%%%%%%%%%%%%%%%%%%%%%%%%%%%%%

\subsection{Future Neutrino Observations}
\label{subsec:neutrino}

IceCube will likely be upgraded to Gen2, which will be much more powerful.  Here we estimate its potential to discover Milky Way sources.

Figure~\ref{fig:nt-gen2} (left panel) shows a scenario where Gen2 does detect Milky Way sources, with the examples of more than one or three sources. As shown, this would strongly constrain the properties of PeVatrons, limiting the parameter space to one of the green areas, as labeled.  The joint parameter space where the LHAASO sources can be 100\% hadronic is even smaller.  The upper right region of the figure (white space) would be ruled out by LHAASO and IceCube not detecting more sources.

Figure~\ref{fig:nt-gen2} (right panel) shows a scenario where Gen2 does not detect Milky Way sources. This would rule out the parameter space where 100\% of LHAASO sources are hadronic and also where $X_{\rm src}$ is larger than $X_{\rm ISM}$. The latter is particularly important, as it would indicate that the total Galactic emission should be dominated by the truly diffuse ISM emission (or leptonic sources). Nevertheless, Gen2 non-detection could still allow a wide range of hadronic models.

The probability that Gen2 will find a source can be viewed optimistically or pessimistically.  If the source parameters are in the band of ``$>$1 Source," there is a good chance that Gen2 finally detects the Milky Way's PeVatrons.  However, even for models that predict one detection, the real source count can easily fluctuate down to zero. On the other hand, if no sources are observed, the region above the dotted line in Fig.~\ref{fig:nt-gen2} (right panel) would be robustly excluded, even in the presence of statistical fluctuations. To put it differently, the left and right panels of Fig.~\ref{fig:nt-gen2} displays prospects for Gen2 in the presence of favorable and unfavorable statistical fluctuations.  Note that sources Gen2 would discover are most likely those already found by LHAASO, but not necessarily.  Gen2, especially with a cascade analysis, can in principle observe a much wider sky region and be sensitive to even extended sources.

Even if Gen2 detects sources, it may not be many, and further improvements in sensitivity will be needed to decisively probe Milky Way hadronic accelerators.  This conclusion is less optimistic than early work that suggested that IceCube could detect several sources~\cite{2006PhRvD..74f3007K, 2007PhRvD..75h3001B, 2008PhRvD..78f3004H, 2009APh....31..437G}.  Those early studies assumed that most gamma-ray sources were hadronic and the calculations were not constrained by detailed measurements of source properties and calibrated to the CR energy budget, as here.  To improve sensitivity to Milky Way sources, {water}-based detectors may be especially important because of their better angular resolution~\cite{2016JPhG...43h4001A, 2018arXiv180810353B,2020NatAs...4..913A,  2022arXiv220704519Y}.  Also, those detectors are planned to be in the Northern hemisphere, and hence would have better sensitivity to the inner Milky Way {through Earth-filtered samples.}

\begin{figure}
    \centering
    \includegraphics[width=\columnwidth]{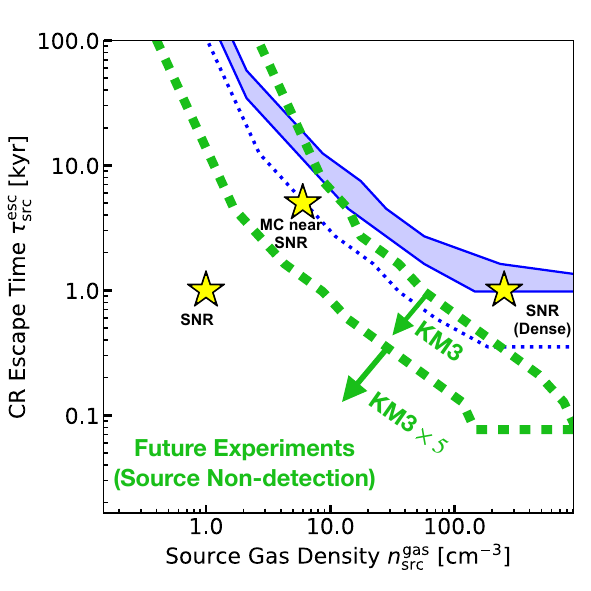}
    \caption{Projected constraints from future neutrino experiments for the case of non-detection. Labels are same as the right panel of Fig.~\ref{fig:nt-gen2} (except for Gen2) and omitted. Some arrows are removed to improve visibility. 
    The neutrino sensitivities are assumed to be uniform across the sky with values as expected for KM3NeT (marked with ``KM3") and five times better than that (``KM3$\times5$"). {\bf \emph{Key takeaway:}} \emph{To definitely probe scenario where hadronic PeVatrons are thin, sensitivity better than that of even Gen2 and KM3NeT is needed.}}
    \label{fig:nt-ocean}
\end{figure}

Figure~\ref{fig:nt-ocean} shows how such future neutrino telescopes could probe PeVatron models. We show the cases of non-detection with two sensitivities (``KM3" and ``KM3$\times$5") as defined above.  Non-detection by ``KM3$\times$5" level experiments would be very constraining. The limit would surpass even the ``MC near SNR" point by a factor of 4, ruling out scenarios where more than quarter of SNRs coincide with MCs that produce gamma rays. Nevertheless, there would still remain a wide range of models allowed, where hadronic PeVatrons are thin, indicating the need for still-better sensitivity.

%%%%%%%%%%%%%%%%%%%%%%%%%%%%%%%%%%%%%%
%%%%%%%%%%%%%%%%%%%%%%%%%%%%%%%%%%%%%%

One idea to attain better sensitivities is further efforts to improve the angular reconstruction of the neutrino cascade events. In the current neutrino source searches, the best sensitivities are obtained by the analysis of track-like events due to muons. As pointed out by Ref.~\cite{2004JCAP...11..009B}, electron neutrinos have much smaller backgrounds than muon neutrinos, making this channel interesting if the angular resolution can be made sufficiently good. We will report on this in a forthcoming paper.

Our analysis does not constrain CR-hidden sources, from which CR do not escape. Though IceCube has not yet identified any such sources, {future neutrino} experiments might.

%%%%%%%%%%%%%%%%%%%%%%%%%%%%%%%%%%%%%%%%%%%%%%%%%%%%%%%%%%%%%%%%%%%%%%%%%%%%%%%%

\subsection{Future Multi-Messenger Observations}
\label{subsec:gamma_etc}

New multi-messenger observations --- covering the full electromagnetic spectrum (not only gamma rays) plus cosmic rays --- will also be important.

The most powerful input we have so far for constraining hadronic PeVatrons is the count of LHAASO gamma-ray sources near the PeV range.  In the next few years, even better results are expected as the LHAASO construction is completed, the observation time is increased, and more theoretical modeling is done.  The LHAASO source count provides firm upper limits on the properties of hadronic PeVatrons, generally stronger than even those based on the non-detection of IceCube neutrino sources.  In the forthcoming years, as Tibet AS$\gamma$, HAWC, and especially LHAASO increase the source counts in the 100~TeV range, this will lead to better constraints in the $n$--$\tau$ plane.  Their diffuse measurements will also help, primarily by constraining the properties of the high-energy cosmic-ray spectrum.

\begin{figure}[t]
    \centering
    \includegraphics[width=\columnwidth]{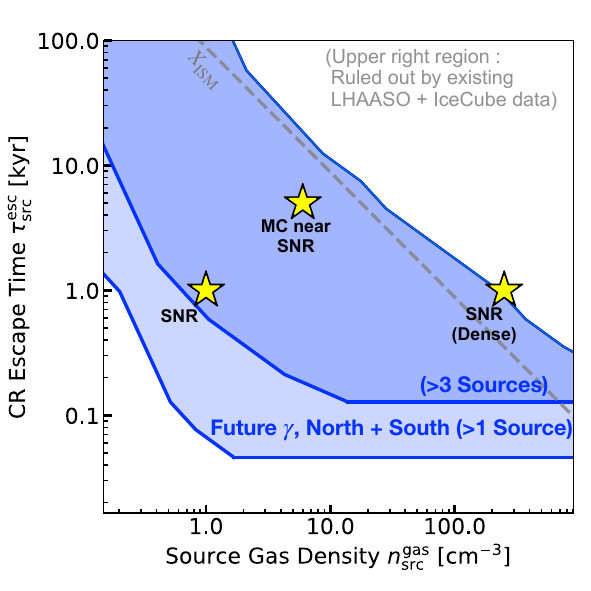}
    \caption{{Sensitivity of SWGO (or ALPACA) + LHAASO to Galactic hadronic PeVatrons. Same as the left panel of Fig.~\ref{fig:nt-gen2}, but we omit constraints from the existing LHAASO and IceCube data for visibility (we instead add a remark in the right top corner). 
    {\bf \emph{Key takeaway:}} \emph{Future gamma-ray observations are promising to find more than one PeVatron, even if $n_{\rm src}^{\rm gas}$ and/or $\tau_{\rm src}^{\rm esc}$ is not large.}
    }}
    \label{fig:nt-sn-future-gamma}
\end{figure}

In the further future, the Cherenkov Telescope Array (CTA)~\cite{2019scta.book.....C}, which will have outstanding flux sensitivity and angular resolution, will discover many sources and will be able to conduct detailed morphological studies, probing the spatial correlations between gamma-ray emission and gas density.  A possible future observatory is the Southern Wide-field Gamma-ray Observatory (SWGO)~\cite{2019arXiv190208429A}.  Like Tibet AS$\gamma$, HAWC, and LHAASO, this would be a large ground array that simultaneously views a wide field on the sky.  Such detectors are more sensitive to extended sources than pointed observatories like CTA.  SWGO would be located in the Southern hemisphere, providing a good view of the inner Galaxy, where a high density of bright, interesting sources is found. Another experiment, the Andes Large-area PArticle detector for Cosmic-ray physics and Astronomy (ALPACA), is also planned to observe highest-energy gamma-ray sources in the Southern hemisphere~\cite{2021ExA....52...85K}. {Figure~\ref{fig:nt-sn-future-gamma} illustrates the power of future gamma-ray observations; we assume that a combination of LHAASO and a southern-hemisphere telesope (like SWGO or ALPACA) will achieve a uniform sensitivity of 0.1~Crab at 100~TeV over the entire sky. Other aspects follow the left panel of Fig.~\ref{fig:nt-gen2}. Even for our regular SNR scenario, more than one detection is expected.}

%%%%%%%%%%%%%%%%%%%%%%%%%%%%%%%%%%%%%%
%%%%%%%%%%%%%%%%%%%%%%%%%%%%%%%%%%%%%%

In principle, gamma-ray observations could separate hadronic and leptonic sources without using neutrinos, though decisive answers have been challenging to obtain.  A lot of attention has been placed on specifically identifying hadronic sources.  This might be done through some combination of the spectrum shape, spatial correlations between emission and gas density, and multi-wavelength observations and modeling.  Another approach would be specifically identifying leptonic sources, using tests of the spectrum shape and multi-wavelength studies.  Crucially, the electrons that produce gamma rays through inverse-Compton emission must also produce synchrotron x-rays.  In the limit that leptonic sources are dominant, this may be a more fruitful approach.  If more sources are identified as leptonic, the allowed parameter space in the $n_{\rm src}^{\rm gas}$--$\tau_{\rm src}^{\rm esc}$ would get narrower and lower, providing stronger constraints on the PeVatron models.

One of the key inputs in our population model is the escape time $\tau_{\rm src}^{\rm esc}$. The recent discovery of ``TeV halos" around pulsars, where the escape time is quite long, has demonstrated the power of gamma rays to probe the escape of very high energy particles from accelerators. Near-future observations will certainly find more halos around CR accelerators, which would be key input to theoretical efforts in understanding the confinement by both leptonic and hadronic sources.

Multi-wavelength observations will help identify sources and characterize their properties. In particular, radio and x-ray observations of gamma-ray objects are important. First, they allow us to isolate components (e.g., SNRs, pulsar wind nebulae, star clusters) due to having good angular resolution. Second, these data probe shock structures and magnetic fields in the region, which is a key input to understand particle acceleration. Third, they are needed to constrain the leptonic component of the gamma-ray emission. In-depth multi-wavelength observations toward sources will also be needed to characterize the environment, constraining the values of $n_{\rm src}^{\rm gas}$.

New CR data will help reduce uncertainties in the CR production rate. First, an improvement in the measurement of the grammage, especially at higher energies, would be crucial in determining the CR energy budget. Second, a better understanding of the CR knee is needed. The existing measurements of CR protons show discrepancies among observations; the amount is by a factor of about 3 at $E_p\simeq3$~PeV, which becomes larger for higher energies. This prohibits us from pointing the exact energy of the proton knee and other spectrum features. Although the all-particle CR spectrum shows the knee at $\simeq 3$~PeV, the proton spectrum might have a spectrum break at $\simeq 1$~PeV. 

Throughout this work, we use the proton energy budget near 1 PeV and also gamma-ray data at $E_\gamma=100$~TeV. In fact, the LHAASO sources are observed much above this energy, reaching more than 1~PeV in some cases. This is curious, because hadronic sources that accelerate protons only to $\sim$1~PeV {would generate gamma rays typically of $E_\gamma \ll E_p$.} Such high-energy photons are also difficult to produce by leptonic sources; in the Klein-Nishina regime of the inverse-Compton scattering, the fractional energy from CRs to gamma rays are large, $E_\gamma/E_e\sim 1$, but the cross section is suppressed. {As the probability distribution of $E_\gamma/E_p$ for hadronic interactions is broad, this suggests that LHAASO are starting to probe the end of radiation spectrum. If LHAASO continues to see sources beyond $\sim$PeV, it might place interesting constraints the locations of proton knee, the contributions from leptonic emission, and the population of super-knee sources.}

%%%%%%%%%%%%%%%%%%%%%%%%%%%%%%%%%%%%%%%%%%%%%%%%%%%%%%%%%%%%%%%%%%%%%%%%%%%%%%%%

\subsection{Importance of Our Population Approach}
\label{subsec:population}

Our new theoretical framework will remain valuable even as studies of individual sources advance.  For example, dedicated IceCube studies of LHAASO sources may lead to stronger constraints on hadronic emission, but those are not connected to the Milky Way CR energy budget.  Even once individual hadronic sources are identified, our population approach will remain valuable.  First, it might be that one or a few hadronic sources are observed in both gamma rays and neutrinos, but that this does not necessarily mean that this source class contributes significantly to the Milky Way's cosmic-ray budget.  In our approach, this can be tested by examining if the source parameters, $n_{\rm src}^{\rm gas}$ and $\tau_{\rm src}^{\rm esc}$, are consistent with the various constraints from gamma rays and neutrinos. Contradictions would suggest that either the sources are not representative PeVatrons or that the standard models of CR production and propagation as adopted above need to be drastically modified.

Our approach can be improved in a number of ways.  Besides the total source count, other observables can be predicted by our population models. First, the flux distributions, $dN/dF$, differ for different combinations of $n_{\rm src}^{\rm gas}$ and $\tau_{\rm src}^{\rm esc}$, even if they predict the same source count. Our calculations here can be regarded as only using the integral of $dN/dF$ above $F_{\rm lim}$. In principle, the use of full distribution can further constrain models. For example, our approach may be allowing too much of the bottom-right region, where sources are luminous and rare.  Second, the latitude distribution could help to constrain models where gamma-ray sources have low gas densities. If the sources are low-luminosity and nearby, then the latitude distribution should have a larger scatter compared to the case of high-luminosity and distant sources. This might further constrain models in the upper left part the plane. Quantitatively, if the {distance} horizon for gamma-ray sources is $\simeq 3\sqrt{n_{\rm src}^{\rm gas}/\rm cm^{-3}}$~kpc and the CR sources are distributed within a height of $\simeq 30$~pc (conservatively small), the scatter in the latitude could be $|b|\simeq0.6^\circ \sqrt{n_{\rm src}^{\rm gas}/\rm cm^{-3}}$.

One difficulty in our approach is that we have to assume a CR source rate to derive constraints in the $n$--$\tau$  plane, because we consider the \emph{number} of sources. Such an assumption would be eliminated if we instead consider the total \emph{flux} from these sources, which we discussed only briefly. Such an approach may give new and comparable constraints on the PeVatron models.  Alternatively, we could construct a plane of $n_{\rm src}^{\rm gas}$ and $\Gamma_{\rm CR}$. Another difficulty in our approach is that we assume that sources are the same as each other, which is a strong assumption, although we briefly discussed how to consider source-to-source variations in an approximate manner.  At the present level of precision, this is reasonable, but ultimately it will be necessary to take variations into account in a more sophisticated manner. 

We encourage the development of public codes to model point sources and collections thereof.  At the moment, there seem to be none.  This is in contrast to extensive work on modeling diffuse gamma-ray and neutrino emission in the Milky Way, such as with GALPROP~\cite{1998ApJ...493..694M, 1998ApJ...509..212S, 2000ApJ...537..763S, 2002ApJ...565..280M, 2004ApJ...613..962S, 2006ApJ...642..902P, 2017ApJ...846...67P, 2019ApJ...887..250P, 2020ApJ...889..167B, 2021arXiv211212745P}, DRAGON~\cite{2014PhRvD..89h3007G, 2017JCAP...02..015E, 2018JCAP...07..006E}, USINE~\cite{2001ApJ...555..585M, 2009A&A...497..991P, 2010A&A...516A..66P, 2020CoPhC.24706942M}, PICARD~\cite{2014APh....55...37K, 2015APh....70...39K}, and CRIPTIC~\cite{2022arXiv220713838K}), some of which are public.

%%%%%%%%%%%%%%%%%%%%%%%%%%%%%%%%%%%%%%%%%%%%%%%%%%%%%%%%%%%%%%%%%%%%%%%%%%%%%%%%
%%%%%%%%%%%%%%%%%%%%%%%%%%%%%%%%%%%%%%%%%%%%%%%%%%%%%%%%%%%%%%%%%%%%%%%%%%%%%%%%

\section{Conclusions}
\label{sec:conclusion}

The Milky Way contains powerful but unidentified accelerators of PeV hadronic CRs.  And gamma-ray point sources have been observed at energies into the PeV range.  However, no neutrino sources have been detected. As illustrated in Fig.~\ref{fig:schematic}, these observations leave two extreme possibilities for the mysterious hadronic PeVatrons: (1) Hadronic CR sources do produce both gamma rays and neutrinos, making them exciting multi-messenger targets, but greater sensitivity is needed to detect the neutrinos, versus (2) Hadronic CR sources are so thin {in matter column density} that they produce no detectable gamma rays or neutrinos {in-situ}, so that the observed gamma-ray sources are only leptonic accelerators.

The main aim of this paper is to understand where the hadronic PeVatrons lie between those extremes.  We introduce a new population-based approach that characterizes sources in the plane of source gas density and CR escape time (the $n$--$\tau$ plane), explicitly calibrating point-source models to CR observations. We quantify the counts of identifiable hadronic sources and the contributions of hadronic sources to the total emission from the Milky Way plane.  We calculate the ranges of allowed models for hadronic accelerators.

We compare predictions for gamma-ray sources to data from LHAASO, which has identified PeV-range sources.  In the optimistic interpretation, these sources could all be hadronic CR accelerators and hence multi-messenger sources.  This requires models to be within the ``100$\%$" band in the $n$--$\tau$ plane shown in Fig.~\ref{fig:nt-current}.  In this case, PeV hadrons accumulate a comparable grammage in the sources and in the ISM.  In a pessimistic interpretation, the LHAASO sources could be all leptonic.  This is allowed over a wide parameter space, quantified in the ``LHAASO (0$\%$ Hadronic)" limit in Fig.~\ref{fig:nt-current} (left).  In this case, the grammage accumulated in the ISM would dominate.  Possibly the most realistic scenario is that only one or a few of the LHAASO sources is hadronic, as in Fig.~\ref{fig:nt-current} (right). 

Neutrino observations are the key to decisively resolving the nature of the hadronic PeVatrons.  We show that IceCube non-detections rule out models with high gas densities, but still leave a wide parameter space open, as per the ``IceCube (Source)" limit in Fig.~\ref{fig:nt-current}.  Improvements in the neutrino sensitivities are needed. {Gen2 and KM3NeT are} promising for finally identifying the hadronic PeVatrons, as discovery can be expected for a wide range of the parameter space (Fig.~\ref{fig:nt-gen2}, left). Non-detections by Gen2 and KM3NeT would constrain significant and important parameter space, but even then, a substantial discovery space for multi-messenger source would remain.  To quantify that, the detection range of Gen2 for is only $\sim 1.3 \sqrt{n_{\rm src}^{\rm gas}/\rm cm^{-3}}$~kpc (for $\Gamma_{\rm CR} = 0.03$~yr$^{-1}$). Even larger detectors may be needed to fully probe the origins of hadronic CRs in the Milky Way.

Identifying the origins of the cosmic rays is a century-old problem.  While it is possible that new observations may soon lead to breakthroughs, it is also possible that this problem will remain challenging for decades more.  A key focus of work on Milky Way hadronic CR accelerators has been on trying to determine if individual gamma-ray sources are hadronic or leptonic, with clear progress but not meeting the goal of definitive answers for most sources.  And even when this goal is achieved, it remains a separate question to decide if these sources are producing enough CRs to account for the Milky Way fluxes.  This paper is a first step in starting a new, population-based approach to revealing the hadronic PeVatrons. Our hope is that it will be improved through new observational constraints, new ideas for theoretical constraints, and the development of public codes similar in sophistication to those used for modeling the Galactic diffuse emission.

In the near term, results from IceCube on observations of the total neutrino emission from the Milky Way plane will be quite important, especially if the hint of a signal strengthens in significance.  As shown in Fig.~\ref{fig:nt-current-diffuse}, an IceCube signal near their present ``Total" limit would suggest that source emission is more important than diffuse emission and that the LHAASO gamma-ray sources should be dominantly hadronic.  Ironically, while that would confirm the existence of hadronic PeVatrons, it would still not reveal where they are.  Still, it would indicate that the optimistic case in Fig.~\ref{fig:schematic} is likely true, indicating a bright future for multi-messenger astronomy as cosmic-ray, gamma-ray, and neutrino observatories gain in sensitivity.  In turn, that would imply excellent prospects not only for resolving long-standing questions in astrophysics, but also for developing new tests of physics beyond the standard model using these sources.

%%%%%%%%%%%%%%%%%%%%%%%%%%%%%%%%%%%%%%%%%%%%%%%%%%%%%%%%%%%%%%%%%%%%%%%%%%%%%%%%
%%%%%%%%%%%%%%%%%%%%%%%%%%%%%%%%%%%%%%%%%%%%%%%%%%%%%%%%%%%%%%%%%%%%%%%%%%%%%%%%

\section*{Acknowledgments}

We are grateful for helpful discussions with Katsuaki Asano, Aya Bamba, Stefano Gabici, Francis Halzen, William Luszczak, Yutaka Ohira, Hidetoshi Sano, Hiromasa Suzuki, and especially Tim Linden and Kohta Murase. {We thank the anonymous referees for their helpful comments.} T.S.\ was primarily supported by an Overseas Research Fellowship from the Japan Society for the Promotion of Science (JSPS).  T.S.\ was partially supported by and J.F.B.\ was fully supported by National Science Foundation grant No.\ PHY-2012955.  This research made use of {\sc matplotlib}~\cite{matplotlib} and {\sc numpy}~\cite{numpy}. Computing resources were provided by the Ohio Supercomputer Center~\cite{OhioSupercomputerCenter1987}. 

%%%%%%%%%%%%%%%%%%%%%%%%%%%%%%%%%%%%%%%%%%%%%%%%%%%%%%%%%%%%%%%%%%%%%%%%%%%%%%%%
%%%%%%%%%%%%%%%%%%%%%%%%%%%%%%%%%%%%%%%%%%%%%%%%%%%%%%%%%%%%%%%%%%%%%%%%%%%%%%%%

\newpage
\clearpage

%%%%%%%%%%%%%%%%%%%%%%%%%%%%%%%%%%%%%%%%%%%%%%%%%%%%%%%%%%%%%%%%%%%%%%%%%%%%%%%%
%%%%%%%%%%%%%%%%%%%%%%%%%%%%%%%%%%%%%%%%%%%%%%%%%%%%%%%%%%%%%%%%%%%%%%%%%%%%%%%%

\clearpage

\appendix

\section{Population Model for Alternative Scenarios}
\label{app:pop-alternative}

\begin{figure}[b]
    \centering
    \includegraphics[width=\columnwidth]{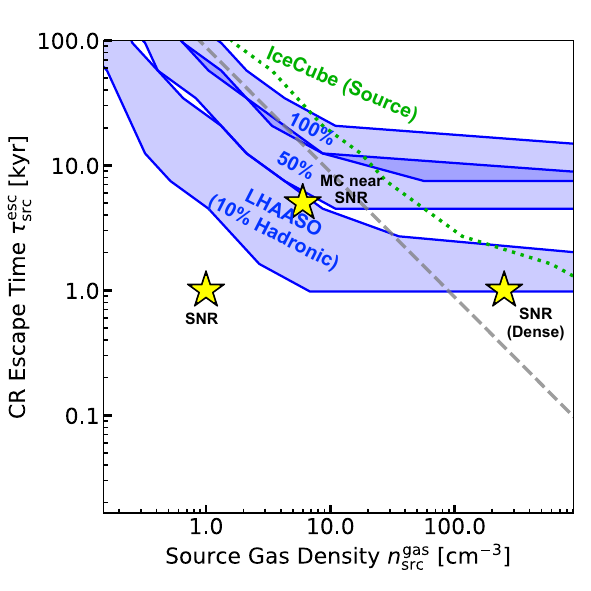}
    \caption{Same as Fig.~\ref{fig:nt-current} (right panel), but for $\Gamma_{\rm CR}$ = 0.003 yr$^{-1}$.}
    \label{fig:nt-rare-2}
\end{figure}

\begin{figure}[t]
    \centering
    \includegraphics[width=\columnwidth]{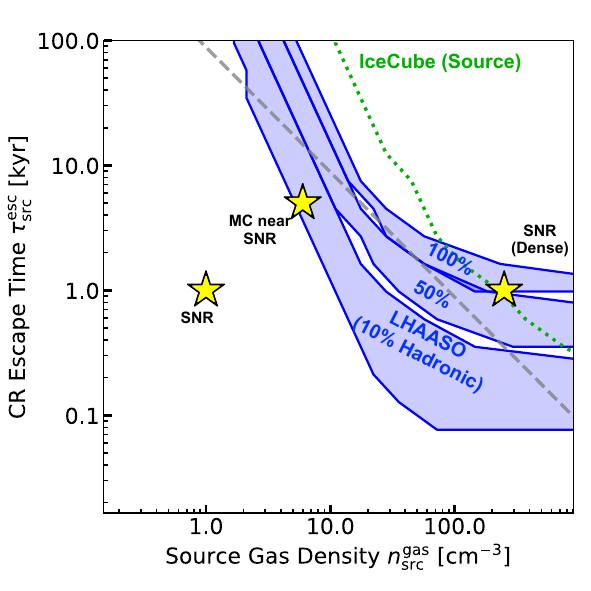}
    \caption{Same as Fig.~\ref{fig:nt-current} (right panel), but for $R_{\rm src} = 30$~pc.}
    \label{fig:nt-30pc-2}
\end{figure}

\begin{figure}[b]
    \centering
    \includegraphics[width=\columnwidth]{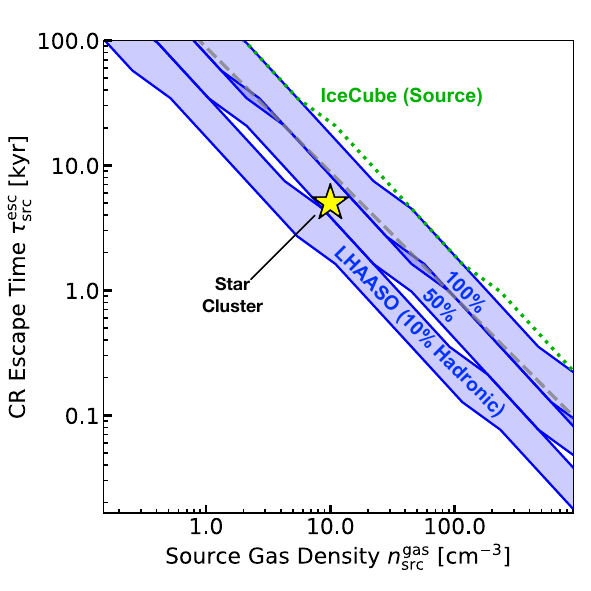}
    \caption{Same as Fig.~\ref{fig:nt-current} (right panel), but for the case of continuous CR injection with $\mathcal{N}_{\rm CR}=10^2$ and $R_{\rm src} = 30$~pc.}
    \label{fig:nt-sc-2}
\end{figure}

\begin{figure}[t]
    \centering
    \includegraphics[width=\columnwidth]{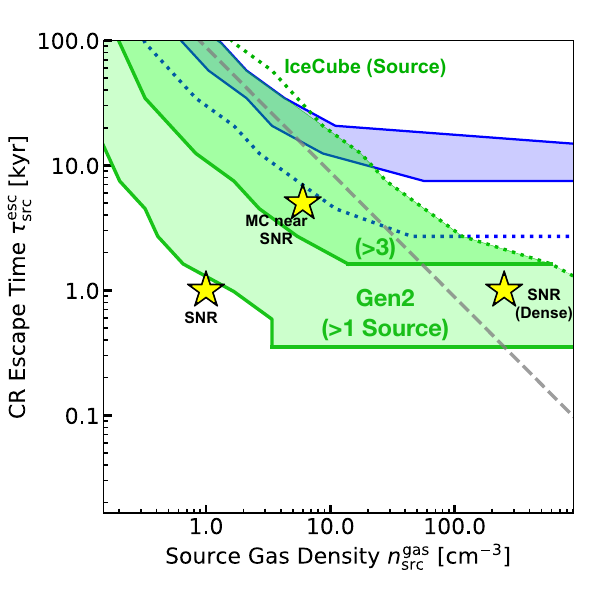}
    \caption{Same as Fig.~\ref{fig:nt-gen2} (right panel), but for $\Gamma_{\rm CR}$ = 0.003 yr$^{-1}$.}
    \label{fig:nt-rare-1}
\end{figure}

\begin{figure}[b]
    \centering
    \includegraphics[width=\columnwidth]{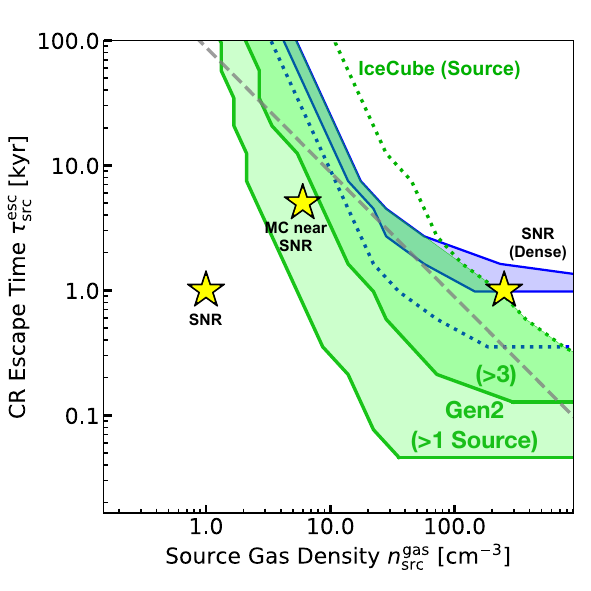}
    \caption{Same Fig.~\ref{fig:nt-gen2} (right panel), but for $R_{\rm src} = 30$~pc.}
    \label{fig:nt-30pc-1}
\end{figure}

\begin{figure}[t]
    \centering
    \includegraphics[width=\columnwidth]{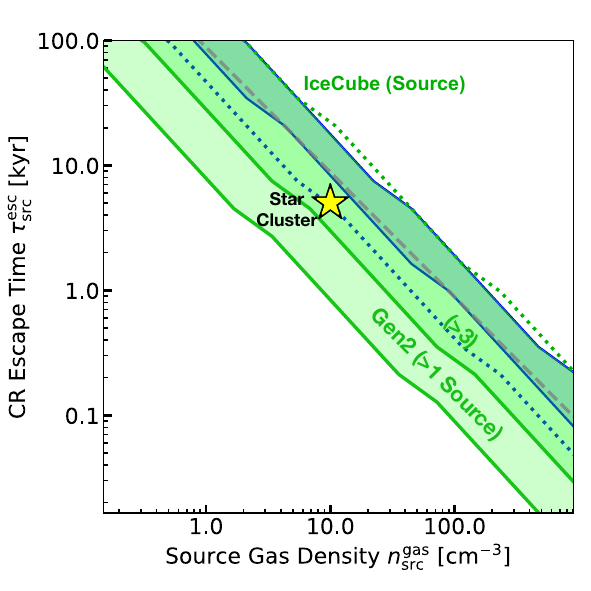}
    \caption{Same as Fig.~\ref{fig:nt-gen2} (right panel), but for the case of continuous CR injection with $\mathcal{N}_{\rm CR}=10^2$ and $R_{\rm src} = 30$~pc.}
    \label{fig:nt-sc-1}
\end{figure}

In the main text, we focused on the case of impulsive CR injections at a rate of $\Gamma_{\rm CR} = 0.03~$yr$^{-1}$ and assumed the size of the gamma-ray and neutrino emitting regions to $R_{\rm src}=10$~pc. Here we present figures for three alternative scenarios:

\begin{enumerate}

    \item Figures~\ref{fig:nt-rare-2} and \ref{fig:nt-rare-1} show scenarios where the CR source rate is 0.003 yr$^{-1}$, 10\% of the Galactic SN rate. 
    
    \item Figures~\ref{fig:nt-30pc-2} and \ref{fig:nt-30pc-1} show scenarios where $R_{\rm src} = 30$~pc. 
    
    \item Figures~\ref{fig:nt-sc-2} and \ref{fig:nt-sc-1} show scenarios where the injections of CRs are continuous. We use $\mathcal{N}_{\rm CR}=10^2$, as discussed in the main text, and we assume $R_{\rm src} = 30$~pc. Also, we show a star to represent the case of ``Star Cluster" sources, as discussed in the main text.

\end{enumerate}

Figures~\ref{fig:nt-rare-2}--\ref{fig:nt-sc-2} are the same as the right panel of Fig.~\ref{fig:nt-current}, and Figs.~\ref{fig:nt-rare-1}--\ref{fig:nt-sc-1} are the same as the left panel of Fig.~\ref{fig:nt-gen2}. In these appendix figures, arrows are omitted for clarity.

%%%%%%%%%%%%%%%%%%%%%%%%%%%%%%%%%%%%%%
%%%%%%%%%%%%%%%%%%%%%%%%%%%%%%%%%%%%%%

%%%%%%%%%%%%%%%%%%%%%%%%%%%%%%%%%%%%%%
%%%%%%%%%%%%%%%%%%%%%%%%%%%%%%%%%%%%%%

%%%%%%%%%%%%%%%%%%%%%%%%%%%%%%%%%%%%%%
%%%%%%%%%%%%%%%%%%%%%%%%%%%%%%%%%%%%%%

%%%%%%%%%%%%%%%%%%%%%%%%%%%%%%%%%%%%%%
%%%%%%%%%%%%%%%%%%%%%%%%%%%%%%%%%%%%%%

%%%%%%%%%%%%%%%%%%%%%%%%%%%%%%%%%%%%%%
%%%%%%%%%%%%%%%%%%%%%%%%%%%%%%%%%%%%%%

%%%%%%%%%%%%%%%%%%%%%%%%%%%%%%%%%%%%%%
%%%%%%%%%%%%%%%%%%%%%%%%%%%%%%%%%%%%%%

%%%%%%%%%%%%%%%%%%%%%%%%%%%%%%%%%%%%%%%%%%%%%%%%%%%%%%%%%%%%%%%%%%%%%%%%%%%%%%%%
%%%%%%%%%%%%%%%%%%%%%%%%%%%%%%%%%%%%%%%%%%%%%%%%%%%%%%%%%%%%%%%%%%%%%%%%%%%%%%%%

\clearpage

\bibliography{icecube_source}

%%%%%%%%%%%%%%%%%%%%%%%%%%%%%%%%%%%%%%%%%%%%%%%%%%%%%%%%%%%%%%%%%%%%%%%%%%%%%%%%
%%%%%%%%%%%%%%%%%%%%%%%%%%%%%%%%%%%%%%%%%%%%%%%%%%%%%%%%%%%%%%%%%%%%%%%%%%%%%%%%

\end{document}